\documentclass[aps,twocolumn,groupedaddress]{revtex4}
\usepackage{graphicx}
\usepackage{CJK}
\usepackage{color}
\usepackage{latexsym}
\usepackage{amsmath}
\usepackage{amsfonts}
\usepackage{amssymb}
\usepackage[autostyle]{csquotes}
\usepackage{bm}
\usepackage{graphicx}
\usepackage{amsbsy}

\usepackage{epstopdf}
\graphicspath{{./figures/}, {../figures/}}

\newcommand{\lla}{\left\langle}
\newcommand{\rra}{\right\rangle}

\begin{document}
\title{Steady state sedimentation of ultrasoft colloids}

\author{Sunil P. Singh}
\affiliation{Indian Institute of Science Education and Research Bhopal, Bhopal by pass road Bhauri, Bhopal 462 066, Madhya Pradesh, INDIA}
\email{spsingh@iiserb.ac.in}

\author{Gerhard Gompper}
\affiliation{Theoretical Soft Mater and Biophysics, Institute for
Advanced Simulation and Institute of Complex Systems,
Forschungszentrum J\"{u}lich, D-52425 J\"{u}lich, Germany}
\email{g.gompper@fz-juelich.de; r.winkler@fz-juelich.de}
\author{ Roland G. Winkler}
\affiliation{Theoretical Soft Mater and Biophysics, Institute for
Advanced Simulation and Institute of Complex Systems,
Forschungszentrum J\"{u}lich, D-52425 J\"{u}lich, Germany}
\email{r.winkler@fz-juelich.de}
\date{\today}

\begin{abstract}
The structural and dynamical properties of ultra-soft colloids---star polymers---exposed to a uniform external force field are analyzed applying the multiparticle collision dynamics approach, a hybrid coarse-grain mesoscale simulation approach, which captures thermal fluctuations and long-range hydrodynamic interactions. In the weak field limit,  the structure of the star polymer is nearly unchanged,  however in an intermediate regime, the radius of gyration decreases, in particular transverse to the sedimentation direction. In the limit of a strong field, the radius of gyration  increases with field strength. Correspondingly, the sedimentation coefficient increases with increasing field strength, passes through a maximum and decreases again at high field strengths. The maximum value depends on the functionality of the star polymer. High field strengths lead to symmetry breaking with trailing, strongly stretched  polymer arms and a compact star polymer body.
In the weak field linear response regime, the  sedimentation coefficient follows the scaling relation of a star polymer in terms of functionality and arm length.
\end{abstract}

\maketitle

\section{Introduction}
External fields are able to induce drastic conformational changes of soft materials, such as polymers, colloids, vesicles etc. In turn, their dynamical and transport properties are modified, an effect which can be exploited in technical applications \cite{Larson_SRF_1999,Bird_DPL_1987}. The understanding of the relation between the nonequilibrium structure and the transport coefficients is fundamental for the rational design of novel functional materials as well as the understanding of the functional principles of biological systems.
The intriguing nonequilibrium properties of soft matter in shear and Poiseuille flow have been illustrated for linear \cite{smit:99,ledu:99,kroe:04,schr:05,gera:06,wink:06.1,jend:04,chel:10,chel:12,stei:12,hara:13} and star polymers \cite{paku:98,Vlassopoulos_MAS_2001,niko:10.1,vlas:14,wink:14.1} as well as  vesicles
\cite{kell82,nogu:04,kant06,misb06,lebe07,vlah07,zhao11,abre14,lamu13} and blood cells
\cite{Abkarian_SSF_2007,
nogu:05,kaou:09,Dupire_DRC_2012,
Pivkin_ACG_2008,McWhirter_FIC_2009,
toma09,Reasor_CLB_2012,
Fedosov_DDC_2014,pelt:13}.

In nature, large macromolecular or colloidal particles sediment to the bottom of a container due to
the gravitational force and the density difference of the particles and the solvent.
Technically, gravity-driven motion is exploited in  analytical ultra-centrifuge techniques  for the characterization and separation of synthetic and biological molecules from mixtures \cite{Harding_PS_1993,Laue_ARBBS_1999}.
Sedimentation of colloidal and polymeric systems is  enormously important for scientific and engineering applications, because soft materials whose size and shape are sensitive to thermal fluctuations and weak external flows,  exhibit interesting, and {\em a priori} unexpected physical behavior. An example is the sedimentation coefficient of DNA molecules in a dilute suspension, which decreases with the increasing driving force \cite{Abbitt_RPB_2003}, denoted as sedimentation anomaly. It is explained by inhomogeneous hydrodynamic interactions of the polymer coil \cite{schl:08,zimm:76,erta:93}.  The coil exterior, especially the chain ends, experience a higher drag, while the monomers in the interior are hydrodynamically shielded. This implies a deformation of the coil and a decreasing sedimentation coefficient \cite{schl:08,schl:07}.
Indeed, the computer simulations of Refs.~\cite{schl:07,schl:08} reveal intriguing conformational changes of the polymer coil with a strong polymer stretching of the trailing end and the formation of a rather compact polymer coil.

In the present work, we investigate the steady-state sedimentation properties of dilute suspensions of the ultrasoft colloids---star polymers. These colloids are particularly interesting due to their intrinsic nature to inhibit colloidal and polymeric properties \cite{stel:00,Vlassopoulos_MAS_2001,vlas:14,wink:14.1}.  A star polymer is a special type of branched polymer, comprised of several  flexible linear polymers which are attached to a common center. The number of polymer arms controls the properties of the colloid---a  small number of arms leads to polymer-like behavior and a large number of arms to colloidal behavior.
The equilibrium and nonequilibrium properties of star polymers have been addressed in various experimental and simulation studies \cite{paku:98,niko:10.1,vlas:14,wink:14.1,gres:87,gres:89,Vlassopoulos_MAS_2001,Ripoll_PRL_2006,sing:12,sing:11,Singh_MACRO_2013,
gupt:12,sabl:17}.

Hydrodynamic interactions are essential for the sedimentation of polymers, as discussed, e.g., in Ref.~\cite{schl:08}. To adequately account for fluid-mediated interactions, we combine molecular dynamics (MD) simulations of a star polymer with the multiparticle collision dynamics (MPC) approach for the fluid \cite{Malevanets_MSM_1999,Kapral_ACP_2008,Gompper_APS_2009}. MPC is a particle based simulation approach, which provides a solution of the Navier-Stokes equations on appropriate length and time scales \cite{male00,huan:12,huan:13}. It includes thermal fluctuations and is excellently suited for a combination with MD simulations \cite{Kapral_ACP_2008,Gompper_APS_2009}. MPC has been shown to provide valuable insight into a broad spectrum of nonequilibrium properties of systems such as polymers \cite{kiku:02,webs:05,ryde:06,fran:08,fran:09,Huang_MAM_2010,chel:10,chel:12}, colloids \cite{padd:04,ripo:06,fedo:12,sing:11,sing:12,niko:10}, vesicles and cells \cite{nogu:04,nogu:05,mcwh:09}, and active particles \cite{tao:10,zoet:12,elge:13,reig:12,thee:14,thee:16.1,yang:14.1,hu:15}.

We find a strong influence of fluid-mediated interactions on the nonequilibrium sedimentation and conformational properties of star polymers. The sedimentation coefficient and the radius of gyration of the star polymer exhibits a non-monotonic behavior. At intermediate field strengths, the coefficient increases with increasing field strength, assumes a maximum and decreases at large field strengths again. Thereby, the increase is more pronounced for star polymers with a larger arm number. The changes in the radius of gyration are strongly linked to those of the sedimentation coefficient, however, with the opposite trend, i.e., the radius of gyration decreases first and increases at large field strengths. Interestingly, the star polymers exhibit a trailing tail at high field strengths, with a few strongly stretched polymer arms.

\begin{figure}
\includegraphics*[width=.48\textwidth]{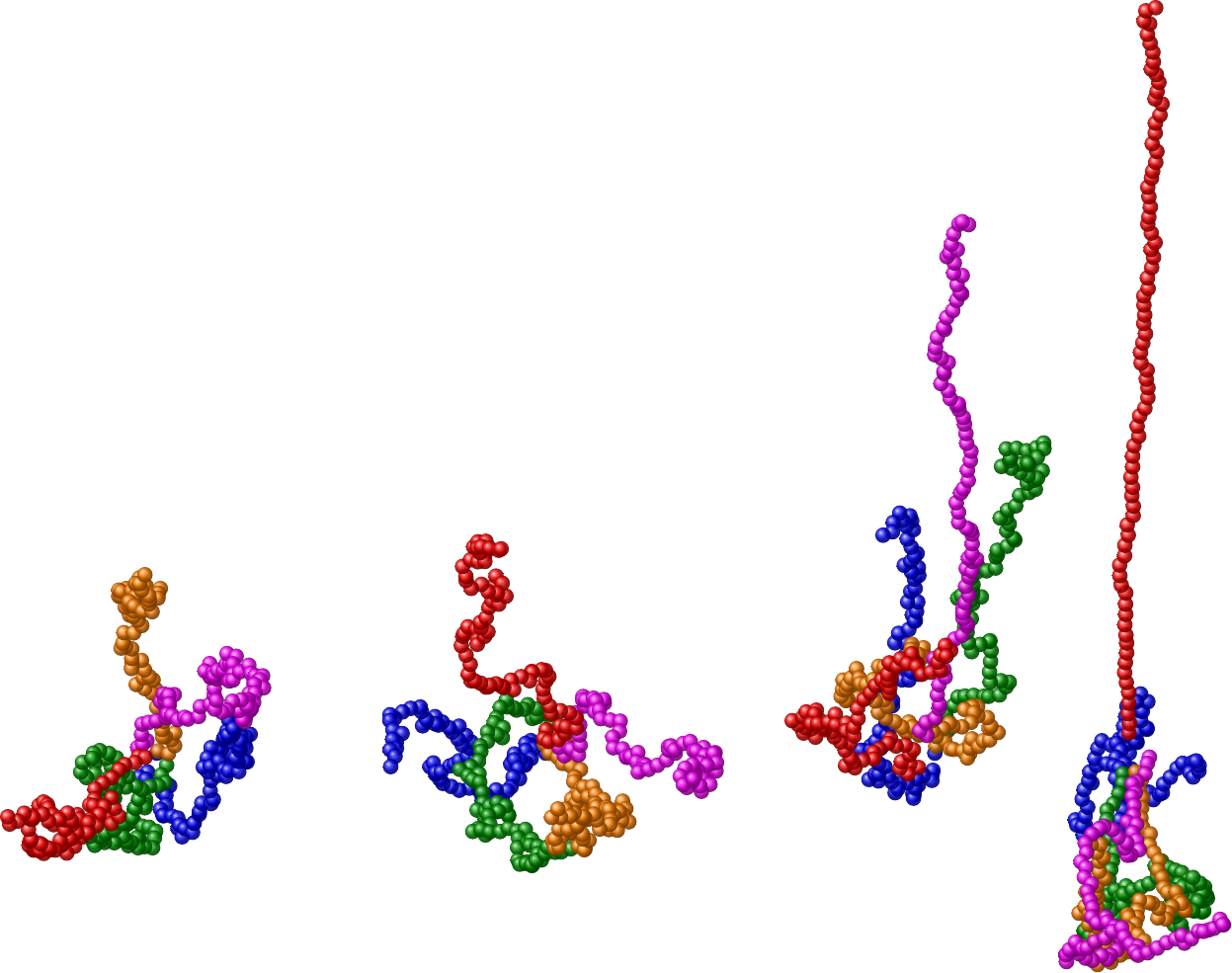}
\caption{Snapshots of a sedimenting star polymer for the arm number $f=5$, arm length $N_m=80$, and several values of the scaled gravitational field strength $G=10^{-3}$, $10^{-2}$, $10^{-1}$,  and $0.5$ (left to right). See also movies in supporting information.}
\label{fig:snap_shot}
\end{figure}

The outline of paper is as follows.  In section \ref{sec:sec2}, the star-polymer model, the coarse-grained description of the explicit solvent, and the interaction of the external field with the polymer are introduced. Section  \ref{sec:sec3} presents results for the conformational and dynamical properties of the star polymers.  All results are summarized and discussed in section\ref{sec:sec4}.

\section{Models} \label{sec:sec2}
\subsection{Star polymer}
We consider a very dilute suspension of star polymers. A star polymer itself consists of $f$ identical flexible linear polymers, which are linked at a common center by one of their ends. A polymer is modeled in a coarse-grained manner as a linear bead-spring chain of $N_m$ beads of mass $M$, hence, the total number of beads are $N_p=fN_m+1$.
The bond potential is given by
\begin{align} \label{eq:pot_bond}
 V_{B} = &  \frac{k_s}{2} \sum_{\mu=1}^{f} \sum_{k=1}^{N_m-1}  \left( |{\bm R}^{\mu}_{k+1} - {\bm R}^{\mu}_{k}| -l \right)^2 \nonumber \\ &  +  \frac{k_s}{2} \sum_{\mu=1}^{f}   \left( |{\bm R}^{\mu}_{1} - {\bm R}_{0}| -l_0 \right)^2,
\end{align}
where ${\bm R}_k^{\mu}$ is the position of monomer $k$ ($k \in \{1,\ldots,N_m\}$) of arm $\mu$ ($\mu \in \{1, \ldots,f\}$),
$l$ is the equilibrium  bond length,  and  $k_s$ denotes the spring constant. The bond length  for the central bead ${\bm R}_0$ is $l_0$. Excluded-volume interactions between non-bonded beads are taken into account by the repulsive,  truncated, and shifted Lennard-Jones (LJ) potential \cite{allen87}
\begin{align} \label{eq:pot_int}
 V_{LJ}= 4 \epsilon \sum_{\nu, \mu = 1}^{f} \sum_{k,j =0}^{N_m} &  \left[ \left(\frac{ \sigma}{ {R}_{kj}^{\nu \mu}}\right)^{12} -
 \left(\frac{\sigma}{{R}_{kj}^{\nu \mu}}\right)^{6}    +\frac{1}{4} \right] \nonumber \\ &  \times \Theta(2^{1/6} \sigma- {\Delta {R}_{kj}^{\nu \mu}}) .
\end{align}

Here, $\Theta(x)$ is the Heaviside step function ($\Theta(x) = 0
~{\rm for}~ x< 0 ~{\rm and}~  \Theta(x) = 1 ~{\rm for}~ x \geq
0$).  The distance between the monomers is $R_{kj}^{\nu \mu} =|{\bm R}_k^{\nu} - {\bm R}_j^{\mu}|$, with ${\bm R}_0^{\nu} \equiv {\bm R}_0$ ($\nu, \mu \in \{1, \ldots, f\}$, $k,j \in \{0, \ldots, N_m\}$). Self-interactions are excluded, i.e., $k\neq j$ for $\nu = \mu$.

Every star-polymer bead is exposed to the gravitational field $\hat {\bm G} = -\hat G {\bm e}_y$, where ${\bm e}_y$ is the unit vector along the direction of the $y$-axis of the Cartesian reference system. Hence, it experiences the force \begin{align}
\bm F_G = M \hat{\bf G} .
\end{align}
In the sedimentation process, fluid is dragged along by a star polymer, which induces a fluid flow. In an experiment, this fluid is reflected by the confining container walls and induces a back-flow. To prevent a net fluid flow in our systems with periodic boundary conditions, we modify the equations of motion of the fluid in such a way that the total momentum of the system (fluid plus star polymer) is zero. By this requirement, fluid back-flow is introduced. This leads to the additional force on a bead
\begin{align} \label{eq:force_back_flow}
{\bm F}_f=-\frac{M^2 N_p}{M N_p + mN_s} \hat{\bm G}  ,
\end{align}
where $m$ is the mass of the fluid particle, $N_s$ is the total number of fluid particles, and $MN_p$ is the total mass of a star polymer.

\subsection{Multiparticle Collision Dynamics}

The ambient fluid is described by the  multiparticle collision dynamics (MPC) approach, an off-lattice, mesoscale, hydrodynamic simulation technique \cite{Malevanets_MSM_1999,Kapral_ACP_2008,Gompper_APS_2009}. In this method, the fluid is represented by point particles with positions ${\bm r}_i$ and velocities ${\bm v}_i$
($i=1,\ldots, N_s$). The particle dynamics proceeds in discrete steps, the streaming  and collision  step. During streaming, the fluid particles of mass $m$ move ballistically in a closed system. However, the gravitational-field induced back-flow has to be taken into account, which yields the velocities and positions after streaming
\begin{align}
\bm v_i(t+h) = & \bm v_i(t) - \frac{M N_p}{M N_p + mN_s}\hat{\bm G}  h  ,  \\
{\bm r}_i(t + h)  = & {\bm r}_i(t) + h {\bm v}_i( t ) - \frac{M N_s}{M N_p + mN_s} \hat{\bm G} \frac{h^2}{2} ,
\end{align}
with the collision time $h$. In the collision step, the simulation box is partitioned into cubic
cells of side length $a$ to define the multiparticle collision
environment. The solvent particles are sorted into these cells and
their relative velocities, with respect to the center-of-mass
velocity of the cell, are rotated around a randomly oriented axis
by an angle $\alpha$, i.e.,
\begin{equation}
\label{eq:collision}
{\bm v}_i(t + h ) = {\bm v}_i(t) + (\mathbf{R} ( \alpha ) - \mbox{\bf I} )
( {\bm v}_i( t ) - {\bm v}_{cm}(t) ) ,
\end{equation}
where $\cal {\bf  R}$ is the rotation matrix, $\mbox{\bf I}$ is the
unit matrix, and ${\bm v}_{cm} =  \sum_{j = 1 }^{N_c } {\bm v}_j /
N_c$ is the center-of-mass velocity of the cell with $N_c$
particles. In this stochastic process, mass, momentum, and energy
are conserved. Momentum conservation  ensures hydrodynamic behavior which emerges on
larger length and time scales \cite{Kapral_ACP_2008,Gompper_APS_2009,huan:12.1}.

The interaction of the star polymers with the fluid is established during
the collision step \cite{muss:05,male00,Gompper_APS_2009,Huang_MAM_2010}. Thereby, the bead velocities are rotated  according to Eq.~(\ref{eq:collision}) similar to those of the fluid particles, with the center-of-mass velocity of the respective collision cell
\begin{align}
 {\bm v}_{cm}(t) = \frac{\sum_{i=1}^{N_c} m {\bm v}_i(t) +
 \sum_{k=1}^{N_c^m} M {\bm V}_k(t) }{m N_c+M N_c^m }  .
\end{align}
Here, $N_c^m$ is the number of beads in the considered cell.
Thereby, momentum is redistributed between fluid and monomers and long-range correlations emerge \cite{huan:13}.

In order to maintain a constant temperature and to remove the energy introduced by the external field, we apply the Maxwell-Boltzmann scaling (MBS) method, which yields a Maxwell-Boltzmann distribution of the fluid-particle velocities  \cite{Huang_JCPS_2010,huan:15}. In the MBS thermostat, the relative velocities---with respect to the center-of-mass velocity of a collision cell---of all particles within such a cell are scaled by a stochastic factor, leaving the dynamical properties of the system unaltered. The stochastic factor is determined from the  Gamma distribution function of the kinetic energy of the particles in a cell.

\subsection{Parameters}

The dynamical behavior of the fluid depends on the various model parameters.
The transport properties of the solvent are determined by the collision
time $h$, the rotation angle $\alpha$, the average number of
particles $\left\langle N_c \right\rangle$ per cell \cite{ihle:03,kiku:03,pool:05,nogu:08,Kapral_ACP_2008,Gompper_APS_2009,wink:09},  which corresponds to the fluid mass density $\rho_s= m \langle N_c \rangle /a^3$.
Small collision times and a large number of average MPC particles
result in fluid-like behavior with a high Schmidt number $Sc$.
In our simulation, we choose parameters such that the transport of momentum due to collision dominates over diffusion. Explicitly, we use the collision time $h/\sqrt{ma^2/(k_B T)}=0.1$, the rotation angle $\alpha =130^\circ$, and $\langle N_c \rangle=10$. These parameters correspond to the
solvent viscosity $\eta_s = 8.7 \sqrt{m k_B T/a^4}$,  kinematic viscosity $\nu_s= \eta_s/\rho_s=0.87 \sqrt{a^2k_B T/m}$, and the
Schmidt number $Sc \approx 17$ \cite{huan:15}.

We study the sedimentation behavior of star polymers  with the  polymer arm lengths $N_m=10, \ 20, \ 40$,  and $80$.
In order to achieve a comparable finite-size effect for the various polymer lengths on the dynamical quantities,
we fix the ratio of the simulation box size along the field direction ($y$-axis) and the radius of gyration of the star polymer for the respective arm length.
A polymer gets elongated in the field  direction, thus the size of the simulation box along the field direction has to be larger than the polymer length.  Explicitly, we apply the following extensions ($L_x, L_y, L_z$) of the simulation box for the various polymer lengths: $N_m=80$,  $L_x/a=80, \ L_y/a =200, \ L_z/a=80$;  $N_m=40$,  $L_x/a=60, \ L_y/a = 130, \ L_z/a=60$;  $N_m=20$ and $N_m=10$,  $L_x/a=40, \ L_y/a = 90, \ L_z/a=40$. Periodic boundary conditions are applied in all spatial directions.  This corresponds to  nearly $10^7$  fluid particles for the polymer length $N_m=80$ and nearly $10^5$ fluid particles for $N_m=10$. In general, $mN \gg M N_s$, hence the correction term for back flow (Eq.~\ref{eq:force_back_flow}) is typically negligible. All the simulations are performed over a range of field strength $G = M \hat Gl /k_BT$, where $10^{-4} \le G < 10^{-1}$.

For the polymer, we use the Lennard-Jones parameters $\epsilon=k_BT$ and  $\sigma/l=0.8$. The parameters for the harmonic bonds are $l=a$ and  $k_s/(k_BT/l^2) =5000$. The mass of a bead
is  $M=10m$. The size of the central bead and the bond lengths to the respective first bead of a polymer arm are twice as large as those of the polymers themselves. This is necessary to allow for a large number of arms to be connected to the central bead.

The velocity Verlet algorithm \cite{allen87} is used to  integrate Newton's equations of motion of the star polymer with the integration time step $h/20$. \\

For an efficient simulation of the system, we apply a hybrid procedure, where a graphics processing unit (GPU) is combined with a CPU. MPC is the most time-consuming part of our simulation. Hence, we divide the computational task into two parts. The
equations of motion of the star polymer are always integrated on the CPU. The MPC dynamics is performed on a GPU. Since MPC streaming of the fluid particles as well as their collisional interactions  are carried out independently, the fluid dynamics is highly parallelizable and can be managed in a efficient way on a GPU.  After every MPC streaming step,  velocities  and positions of the monomers are transferred from the CPU to the GPU for the collisional interaction with the  fluid particles. After the collision with fluid velocities are transferred back to the CPU for the integration of the bead equations of the solute. A detailed description of the GPU implementation of  MPC is provided in Ref.~\cite{west:14}.

\begin{figure}
\includegraphics*[width=\columnwidth]{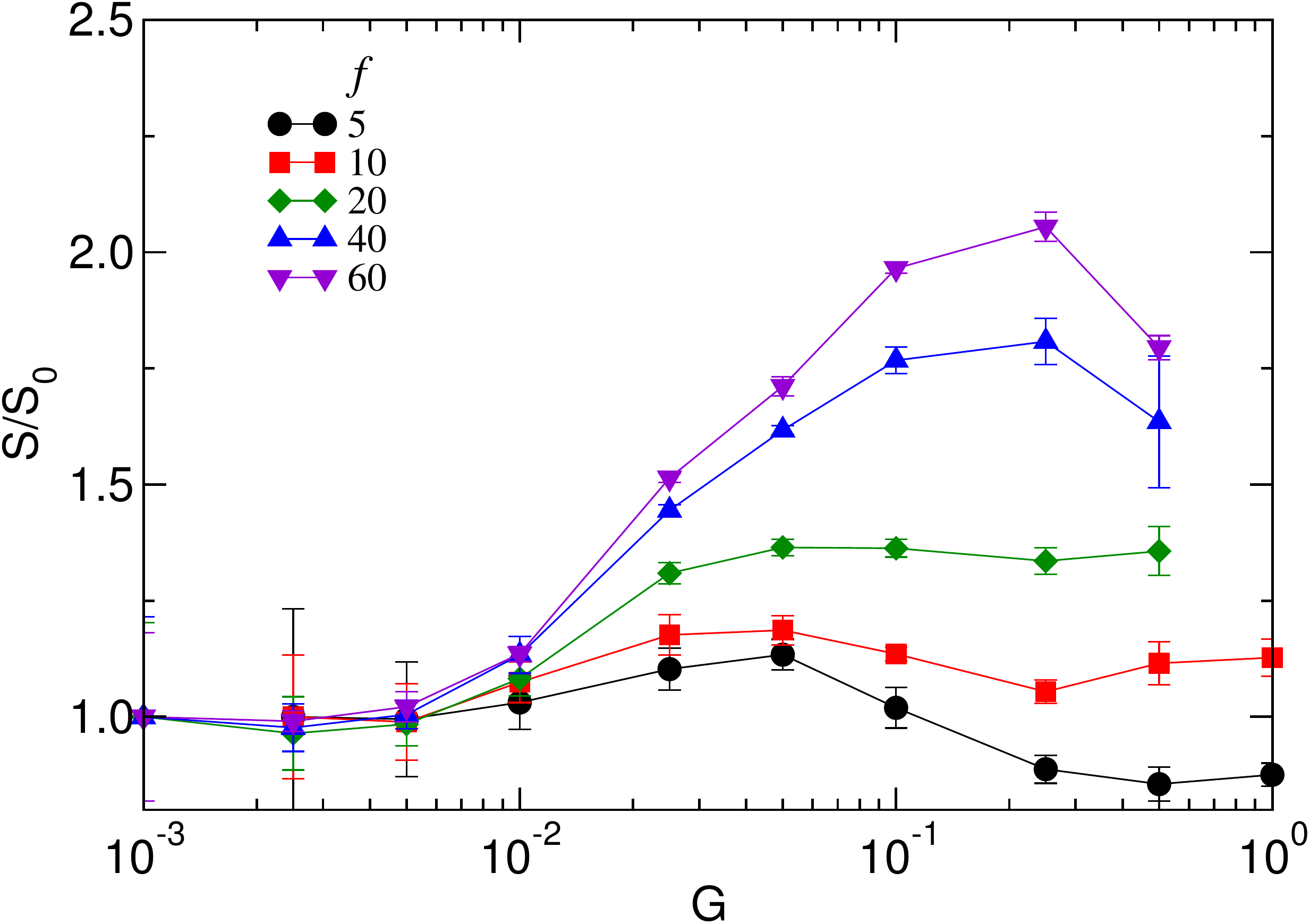}
\caption{Normalized sedimentation coefficients $S /S_0$ of  star polymers with the indicated functionalities
for the arm length $N_m=80$ as a function of the external field $G = M \hat G l /(k_BT)$. $S_0$ is the sedimentation coefficient of the asymptotic weak-field limit.}
\label{fig:S_g}
\end{figure}

\section{Results} \label{sec:sec3}

\subsection{Sedimentation Coefficient}

Under the influence of the external field and in the stationary state, the star  polymer drifts along the direction of the field, with a constant average velocity. Hence, the magnitude of the total external force $F_T=MN_p \hat G$ is equal to magnitude of the frictional force, i.e.,
\begin{equation}
M N_p \hat G = \gamma V_{cm} ,
\end{equation}
where $\gamma$ is the total friction coefficient of the star polymer  and ${\bm V}_{cm} = \sum_{i=1}^{N_s} \lla{\bm V}_i \rra/N_s$ its center-of-mass velocity. The ratio of the center-of-mass velocity and the external force defines the sedimentation coefficient $S$, thus
\begin{align}
S \equiv \frac{V_{cm}}{\hat G} = \frac{M N_p}{\gamma} .
\end{align}

Using Stokes relation, the friction coefficient ($\gamma=6 \pi \eta R_h$) is proportional to the hydrodynamic radius $R_h$ of the star polymer. With the approximation  of the hydrodynamic radius by the radius of gyration $R_g$, for which scaling arguments yield the relation
\begin{align} \label{eq:gyration_radius}
 R_g \sim l N_m^{\nu} f^{(1-\nu)/2},
\end{align}
with the critical exponent $\nu \approx 0.6$ \cite{daou82,birsh86,gres:87}, the sedimentation coefficient should exhibit the scaling relation
\begin{align} \label{eq:sedimentation_scaling}
S \sim N_m^{1-\nu} f^{(1+\nu)/2}
\end{align}
at least for unperturbed star polymers at low external forces. Our simulation studies of Ref.~\cite{sing:14} on the diffusive dynamics of star polymers of various functionalities confirm approximately the dependence $R_g \sim f^{(1-\nu)/2}$, with  $\nu \approx 0.63$ for the considered short polymers, but show a somewhat stronger dependence of the hydrodynamic radius on $f$, namely $R_h \sim f^{\hat \delta}$ with $\hat \delta = 0.29$ instead of $0.2$. We like to mention that in the free draining limit, the friction coefficient is proportional to $N_m f$ and the sedimentation coefficient is independent of the star molecular weight.

Figure~\ref{fig:S_g} displays sedimentation coefficients for various functionalities as function of the scaled strength $G=M \hat Gl/k_BT$ of the external field. Note, $l$ corresponds to the Kuhn length of the polymer. The curves are normalized by the respective asymptotic sedimentation coefficient $S_0$ in the limit of vanishing field. As expected, the sedimentation coefficient is independent of $\hat G$ in the linear response regime for all functionalities. To achieve accurate results,  we have generated nearly 50 independent data sets for every $f$ in the weak-field limit, because here thermal fluctuations are strong and the drift is weak. In an intermediate regime, $S/S_0$ increases with increasing $G$, passes through a maximum and decreases again.
Since we are limited in the range of applicable forces, we cannot extent our studies to large $G$ and, hence, cannot comment on the behavior for asymptotically large values. However, we observe a strong dependence on the functionality. Thereby, $S/S_0$ increases with increasing $f$ for intermediate field strengths and the maximum shifts to larger $G$. This is certainly related to considerable conformational changes of the star polymer as illustrated in Fig.~\ref{fig:snap_shot}.

\begin{figure}[t!]
\includegraphics*[width=\columnwidth]{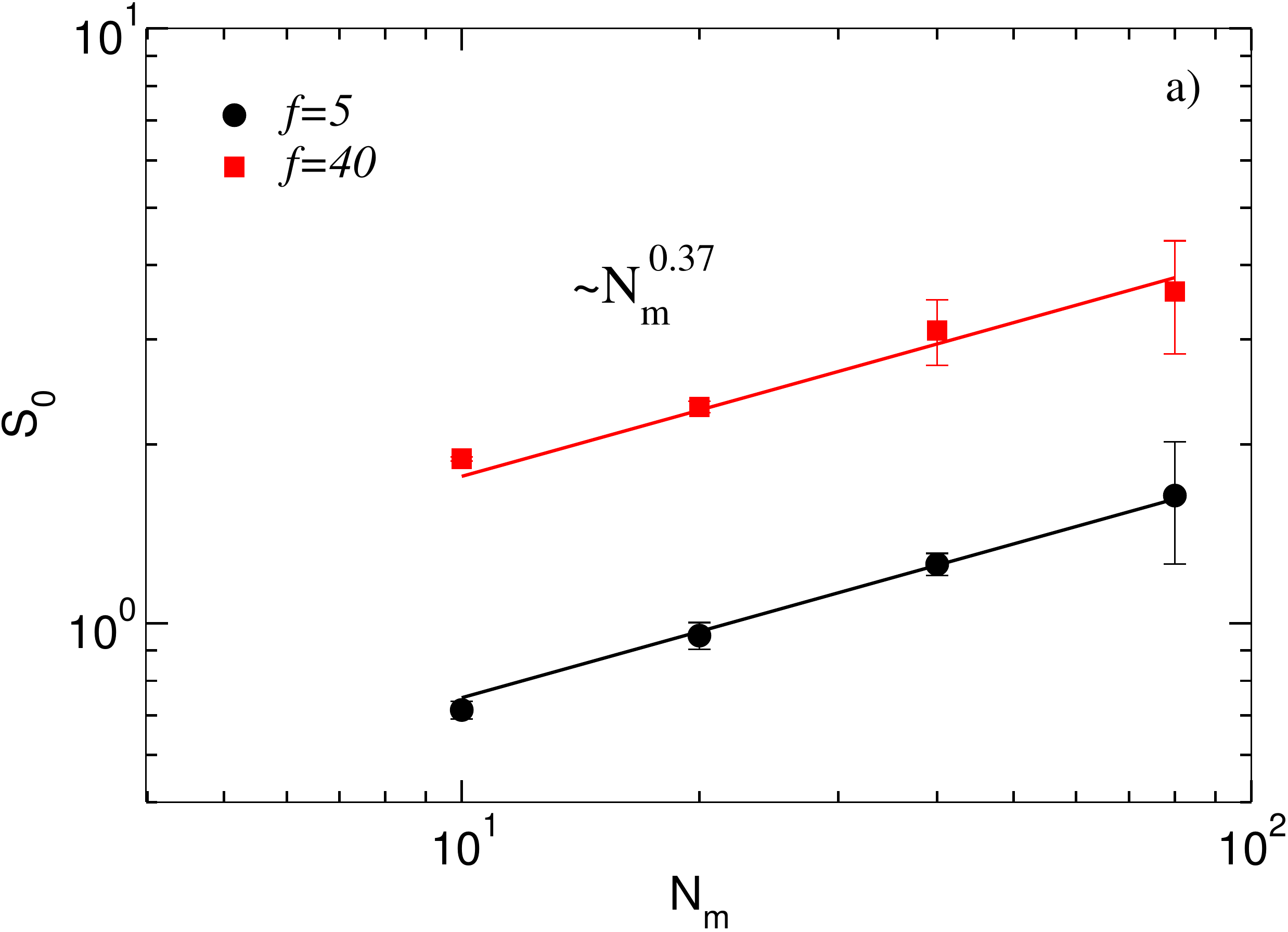}
\includegraphics*[width=\columnwidth]{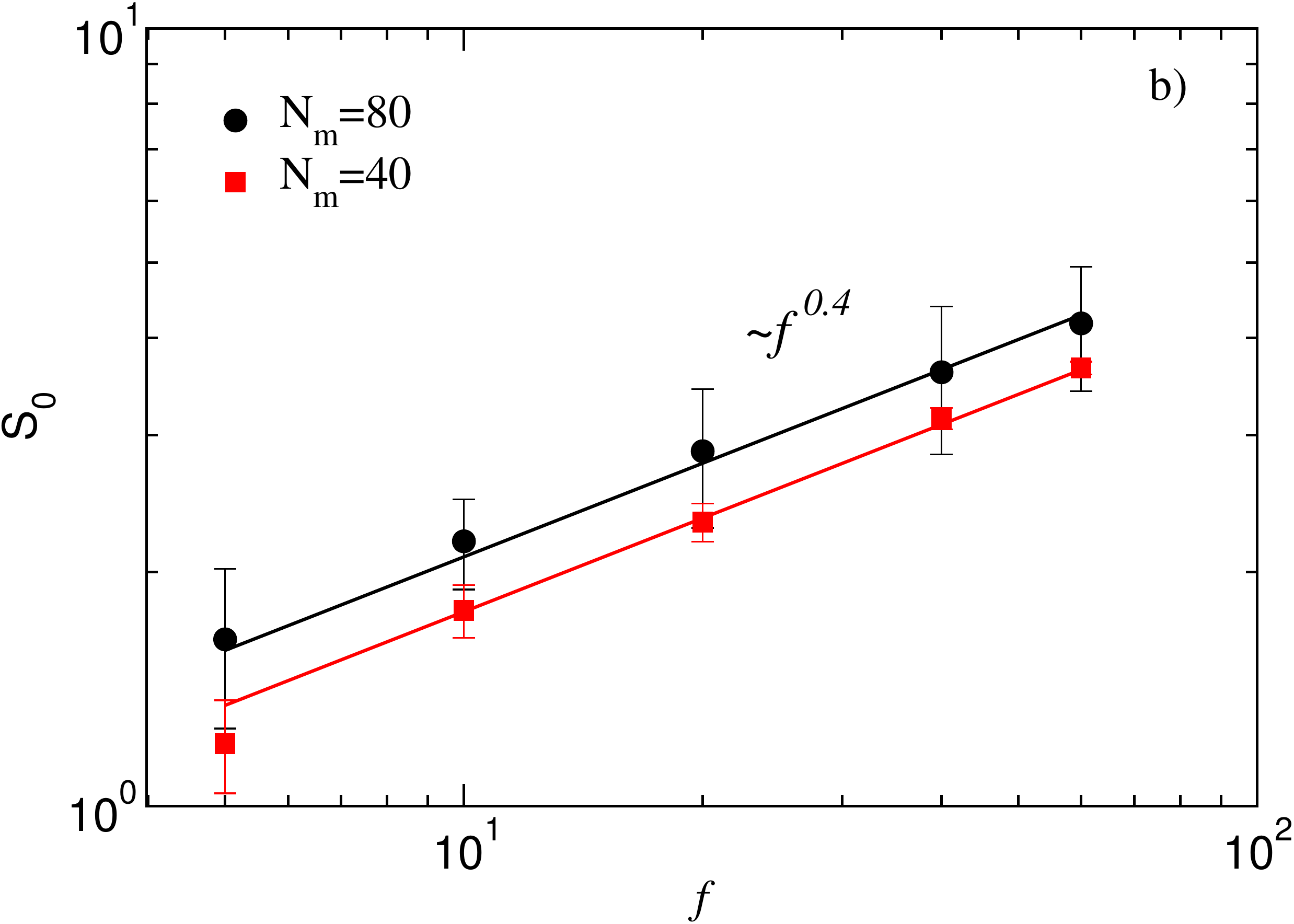}
\caption{a) Dependence of the sedimentation coefficient $S_0$  on the length of polymer arms for
$f=5$ and $f=40$.  The solid lines illustrate the power-law dependence $S_0 \sim N_m^{1-\nu}$, where $\nu \approx 0.63$. b) Dependence of the sedimentation coefficient $S_0$ on the number of polymer arms $f$ for the arm lengths  $N_m=40$ and $N_m=80$. The solid lines are the power-law fits $S_0\sim f^{\delta}$, with $\delta \approx 0.4$.}
\label{fig:S_arm_g}
\end{figure}

The dependence of the sedimentation coefficient $S_0$ on the arm length and number is shown in Fig.~\ref{fig:S_arm_g}.  The values of $S_0$ for various $N_m$ and $f$ are obtained in the linear response regime in the weak field limit. In this regime, $S_0$ is independent of $G$.
 In accord with the scaling prediction of Eq.~(\ref{eq:sedimentation_scaling}), $S_0$ increases with increasing $N_m$ and $f$, respectively.
Thereby, we find $S_0 \sim N_m^{0.37}$ in close agreement with Eq.~(\ref{eq:sedimentation_scaling}) for $\nu\approx 0.63$. The latter value is consistent with various simulation studies of equilibrium and nonequilibrium properties of star polymers  for the considered arm lengths \cite{fedo:12,sing:12,Singh_MACRO_2013,sing:14}. For the dependence of $S_0$ in $f$, we find the power law $S_0 \sim f^{\delta}$, with $\delta \approx 0.4$ independent of polymer length. However, $\delta$ is significantly smaller than the value predicted by scaling considerations [Eq.~(\ref{eq:sedimentation_scaling})], which is $0.82$. Even if we consider the somewhat stronger dependence $R_h \sim f^{0.29}$ on functionality, the value $\delta = 0.4$ is significantly smaller than the theoretical prediction. The origin of the discrepancy remains to be resolved, but back flow might influence the hydrodynamic interactions between the beads.

\begin{figure}[t!]
\includegraphics*[width=\columnwidth]{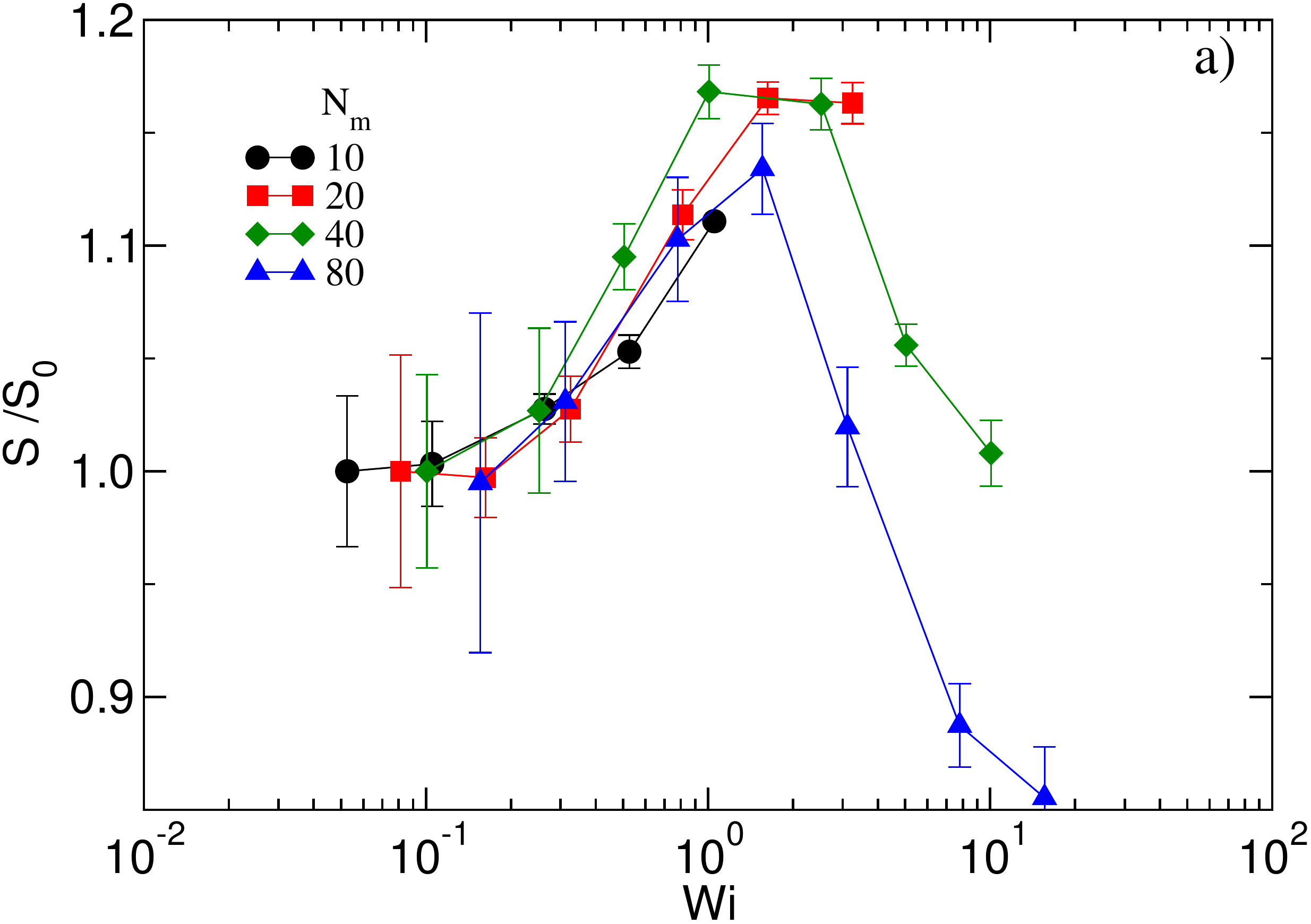}
\includegraphics*[width=\columnwidth]{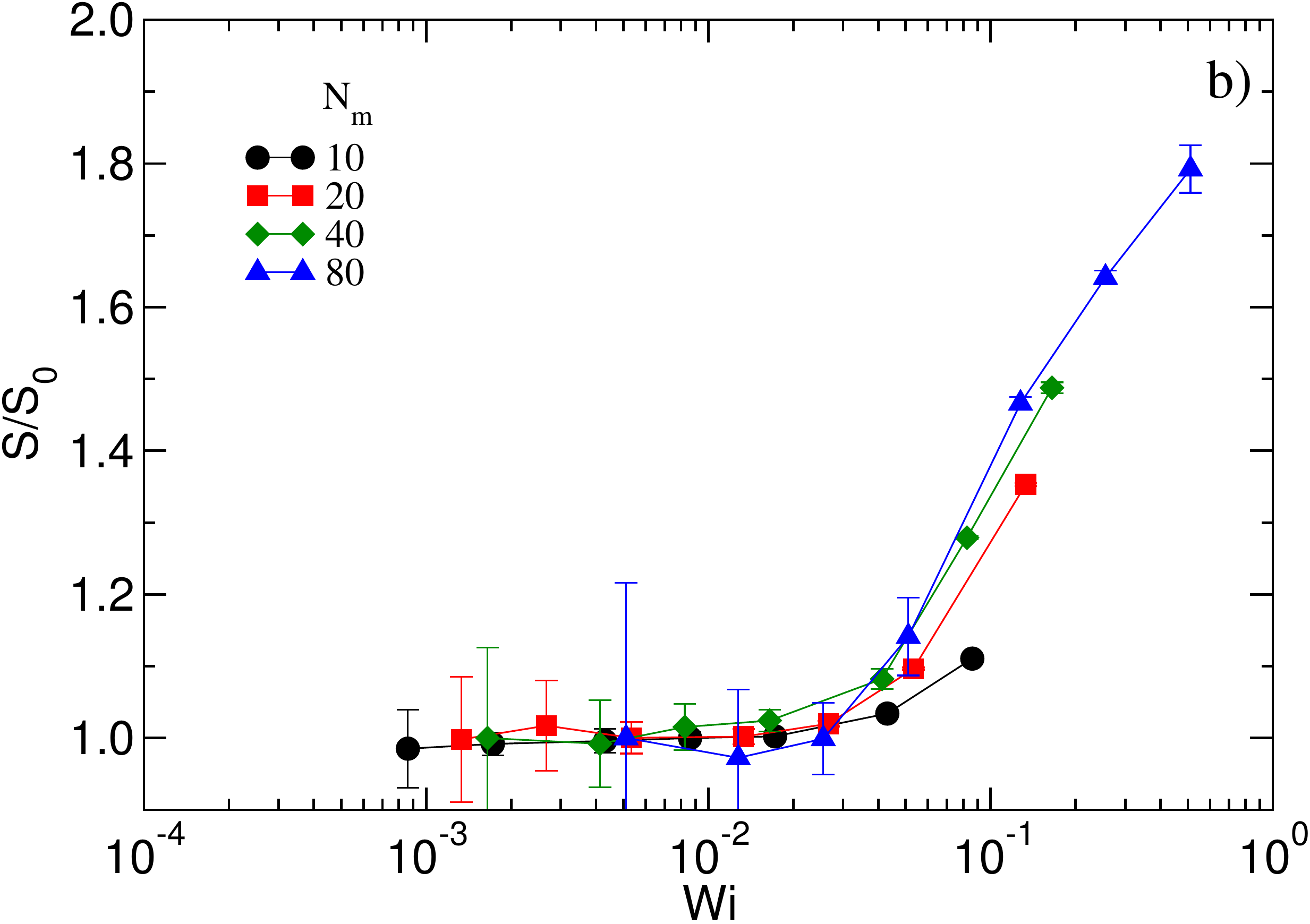}
\caption{Scaled sedimentation coefficients $S/S_0$ of star polymers  with the indicated arm lengths  as function of the Weissenberg number $Wi$ (Eq.~(\ref{eq:weissenberg})) for the arm numbers a) $f=5$  and b) $40$.}
\label{fig:S_arm_40}
\end{figure}

Figure \ref{fig:S_arm_40} displays scaled sedimentation coefficient $S/S_0$ for the polymer lengths $N_m=10,\ 20, \ 40$, and $80$, and the two different functionalities $f=5$ and $40$ as function of the Weissenberg number $Wi$.
The Weissenberg number is defined as follows. At weak  external fields, a star polymer experiences a shear force on its surface during sedimentation, which gives rise to the shear rate $\dot \gamma \sim V_{cm}/R_g$. Within the blob model of a star polymer \cite{gres:87,gres:89}, this leads to the scaling relation for $\dot \gamma$ in terms of the arm length and functionality
\begin{align}
\dot \gamma \sim \hat G N_m^{1-2\nu} f^{\nu} .
\end{align}
The relaxation of a polymer arm is dominated by the relaxation of its largest blob \cite{gres:89} and, hence, we define a Weissenberg number via $Wi=\dot \gamma \tau_B$, with the blob relaxation time $\tau_B$. In the presence of hydrodynamic interactions, $\tau_B \sim R_B^3$, where $R_B$ is the blob radius. The latter scales as $R_B\sim R_gf^{-1/2} \sim N_m^{\nu} f^{-1/2}$ with arm length and functionality \cite{gres:89}. Thus, we finally obtain the scaling relation for the Weissenberg number
\begin{align} \label{eq:weissenberg}
Wi \sim \hat G N_m^{1+\nu} f^{\nu-3/2} .
\end{align}
 In the following, we present the sedimentation coefficient as function of the Weissenberg number, taking  $Wi$ as $Wi=G N_m^{1+\nu} f^{\nu-3/2}$.
As displayed in Fig.~\ref{fig:S_arm_40}, a reasonable scaling of the curves for various arm lengths is only achieved for the functionality $f=40$ and longer arms. The predicted dependence on functionality is not reproduced by the simulations. This is not surprising, since the obtained scaling in Fig.~\ref{fig:S_arm_g} b) deviates from the simple scaling prediction.

As with increasing functionality, the ratio of $S/S_0$ increases with increasing external field strength in the non-linear response regime. This increase is consistent with simulation results of linear polymers \cite{schl:07}. The onset of the non-linear regime is reasonably well captured by the prediction (\ref{eq:weissenberg}).
As displayed in Fig.~\ref{fig:S_arm_40} b), in particular for the longer polymer arms, the  sedimentation coefficient shows a non-monotonic dependence on the external field; it passes through a maximum value and then decreases again with increasing $Wi$. We expect a similar behavior for the shorter polymers, however, for them, we cannot reach large field strengths without violating limitations of the MPC method, e.g., small Mach numbers. The influence of the external field seems to be more pronounced for star polymers with a larger number of arms. Over the accessible range of field strengths, the values $S/S_0$ for star polymers of functionality $f=40$ are always higher than their linear-response-regime values (Fig.~\ref{fig:S_arm_40}b)), whereas the values of $S/S_0$ for $f=5$ are below the linear-response-regime values (Fig.~\ref{fig:S_arm_40}a)).

\begin{figure}[t!]
\includegraphics*[width=\columnwidth]{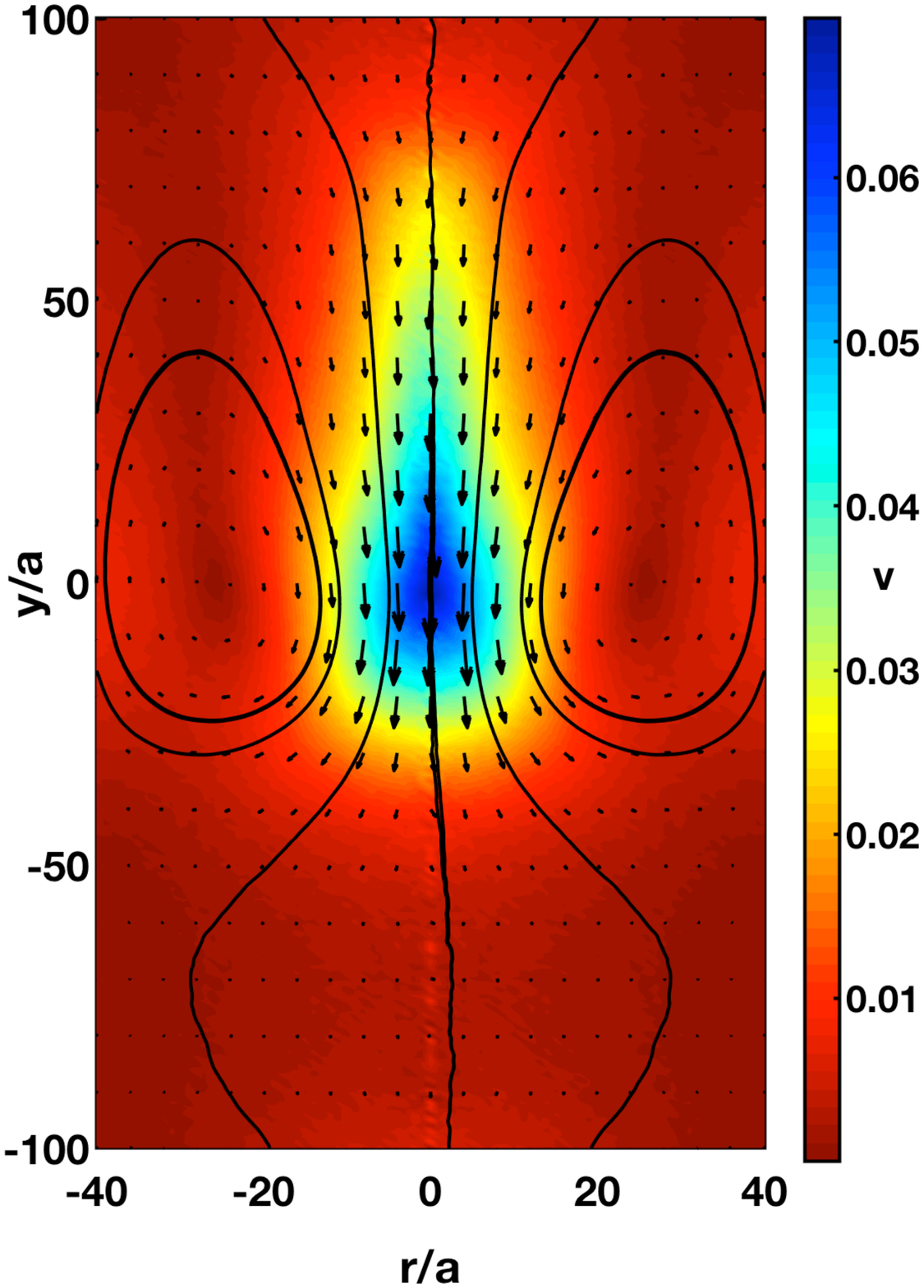}
\caption{Flow field of a star polymer in the laboratory reference frame for the arm length and number $N_m=80$, $f=60$, and field strength $G=0.5$. The coordinate $r$ is the radial distance with respect to sedimentation direction ($y$ axis). The star polymer drags along fluid, which is indicated by the velocity-field vectors and solid black line.  The flow lines  in the head region reflect the recirculation of fluid \cite{schl:08}. Despite the strong field corresponding to Reynolds numbers larger than unity, the flow is laminar. }
\label{fig:flow_profile}
\end{figure}

Considering the sedimentation velocities and the sizes of the star polymers, a remark on the Reynolds number is in order. Taking characteristic values for the sedimentation velocity and the  radius of gyration, the Reynolds number $Re$ is $Re = 2 S_0 \hat G R_{g0}/\nu \approx 10 G$ for $S_0=3$ and $R_{g0} =15a$. Hence, the Reynolds number is larger than unity for $G \gtrsim 0.1$. This implies that the observed saturation or weak decrease of the sedimentation coefficient (Fig.~\ref{fig:S_g}) appears for Reynolds numbers larger than unity. A priori, the effect of the Reynolds number on $S$ in this regime is not evident. A comparison of our results with those of Refs.~\cite{schl:08,schl:07} at zero Reynolds number for linear and ring polymers shows qualitative agreement---$S$ increases first with increasing $G$ and decreases again at larger $G$. Thereby, in Refs.~\cite{schl:08,schl:07} larger $G$ values are considered. Hence, from a qualitative point of view, we consider our results for star polymers as representative and expect a similar behavior for smaller Reynolds numbers. This is supported by Fig.~\ref{fig:flow_profile}, showing a non-turbulent flow field of a sedimenting star even at $Re>1$. In any case, the simulation results for the larger $G$ values reflect the sedimentation behavior of  star polymers at the respective Reynolds numbers.

\subsection{Structural Properties}

Strong external fields induce large-scale conformational changes of the ultra-soft colloids, as illustrated in Fig.~\ref{fig:snap_shot} for our star polymers.  In-order to characterize these conformational changes,  we compute the radius-of-gyration tensor, which is defined as
\begin{equation}
 G_{\alpha \beta} = \frac{1}{N_s} \left\langle \sum_{i=1}^{N_s}
 \Delta R_{i\alpha}\Delta R_{j \beta} \right\rangle .
\end{equation}
Here,  $\Delta R_{i\alpha }$ is the position of the $i^{th}$ beat relative to  the star center-of-mass, and $\alpha$, $\beta \in\{x,y,z\}$.
In the limit of a vanishing field, a star polymer is isotropic and all the diagonal components  are equal, i.e.,
$G_{\alpha  \alpha} = G^{00}_{\alpha \alpha} = R^2_{g0}/3$, where $R_{g0}$ is the equilibrium radius of gyration [Eq.~(\ref{eq:gyration_radius})].

\begin{figure}[t!]
\includegraphics*[width=\columnwidth]{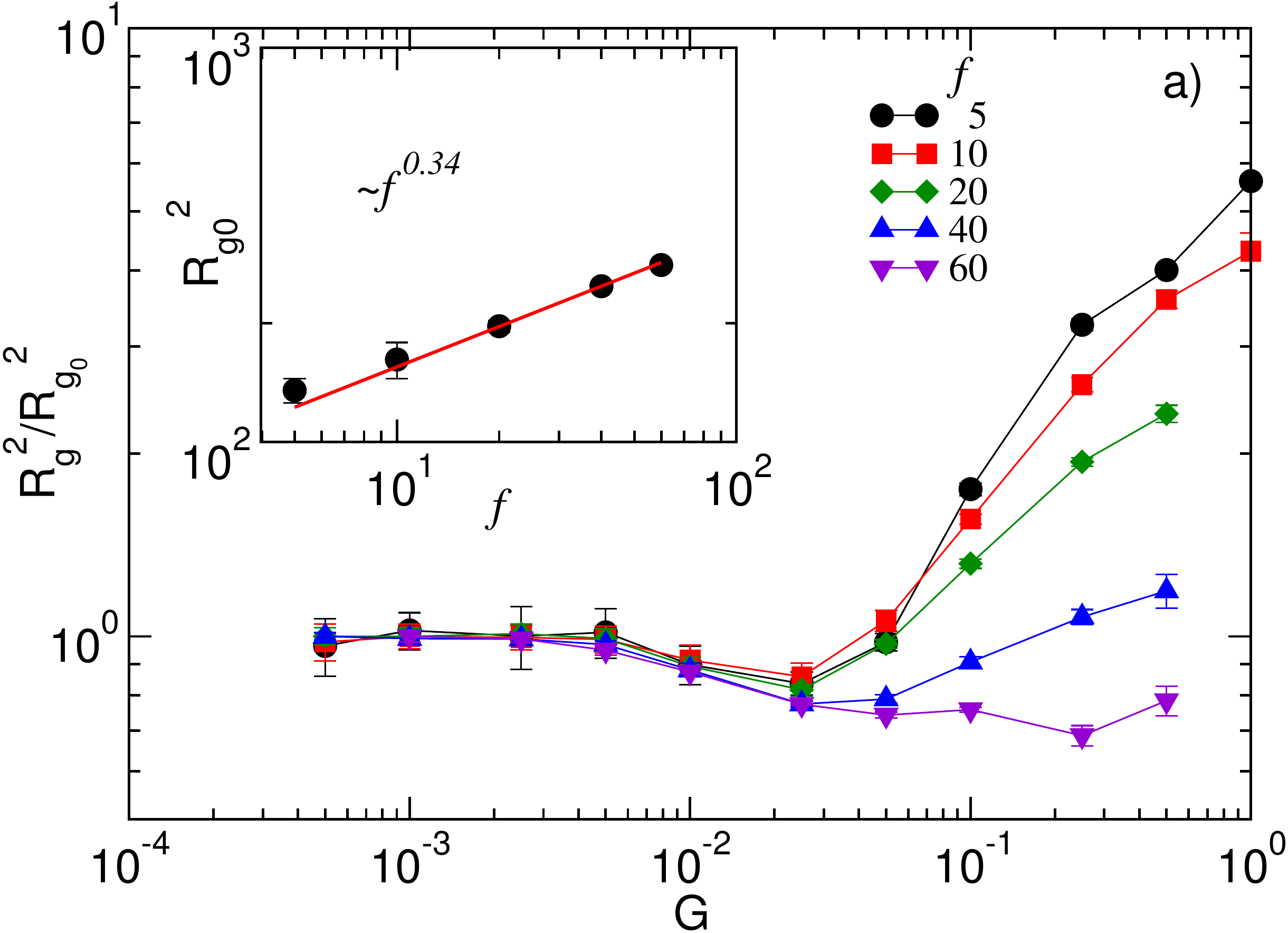}
\includegraphics*[width=\columnwidth]{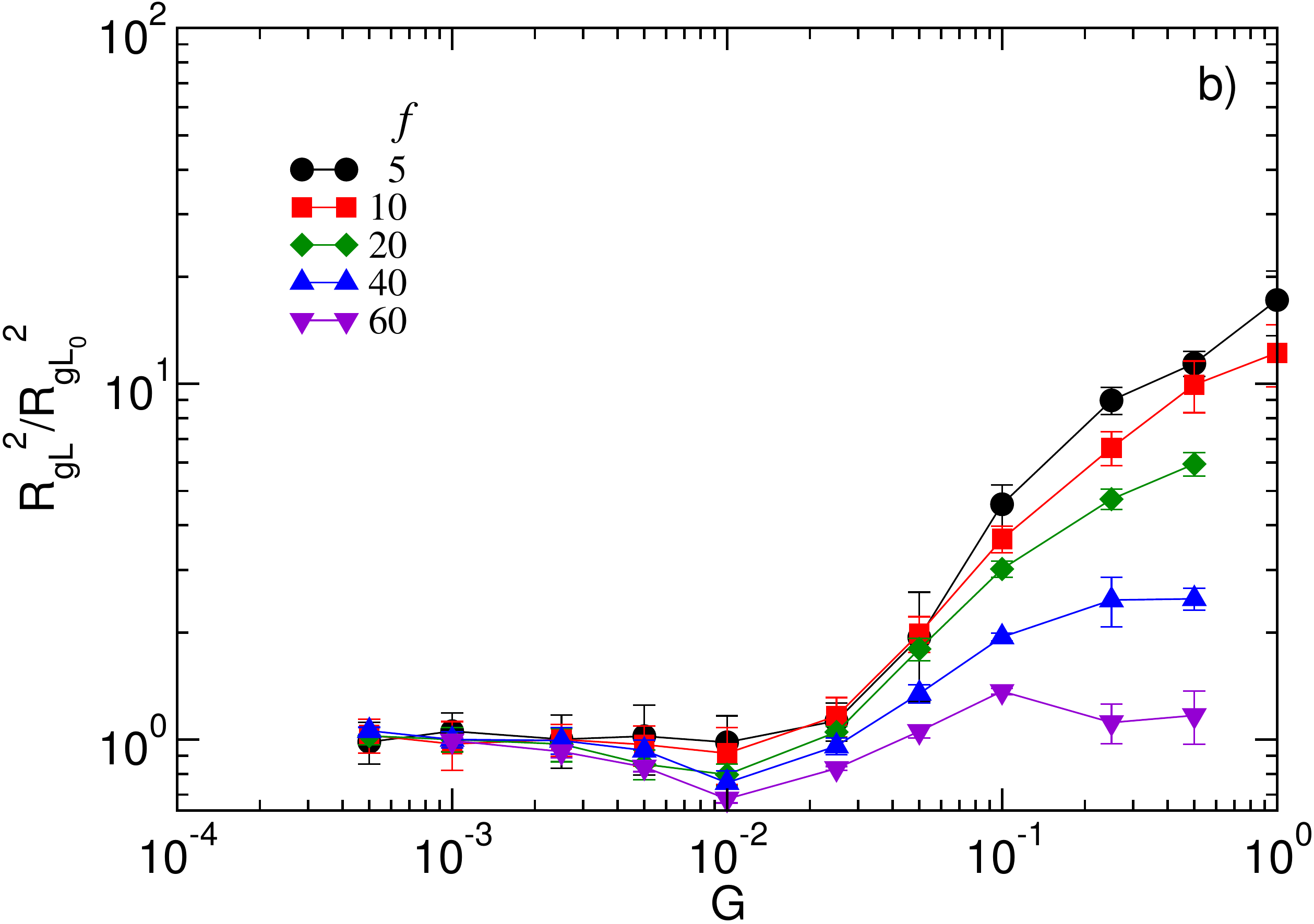}
\includegraphics*[width=\columnwidth]{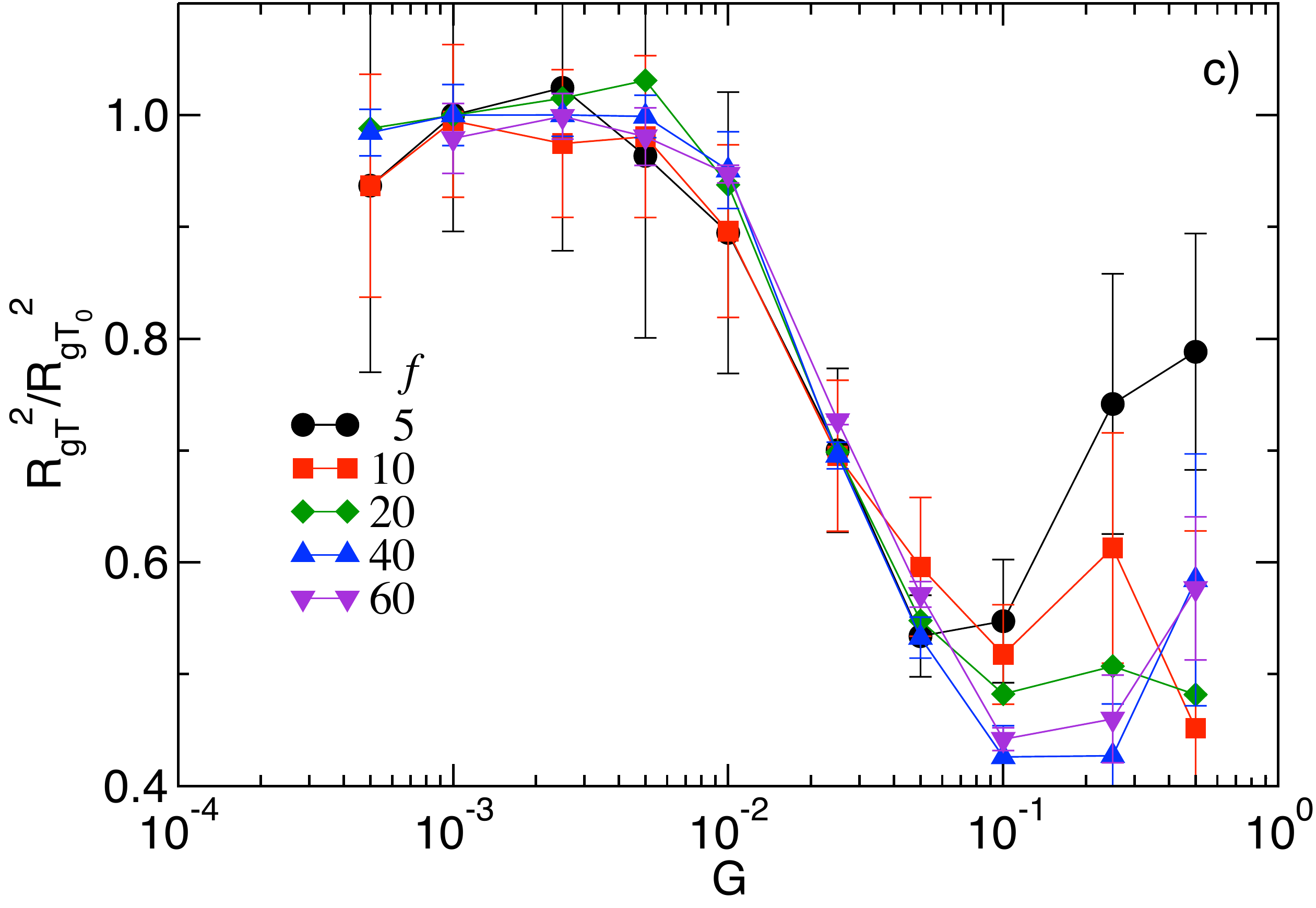}
\caption{a) Overall radii of gyration of star polymers and their components b) along and c) perpendicular to the field as function of the external field for the indicated functionalities and the arm length  $N_m=80$. The  inset in a) shows the dependence of the equilibrium radius of gyration $R_{g0}^2$ on the functionality $f$. }
\label{fig:rg}
\end{figure}

\begin{figure}[t!]
\includegraphics*[width=\columnwidth]{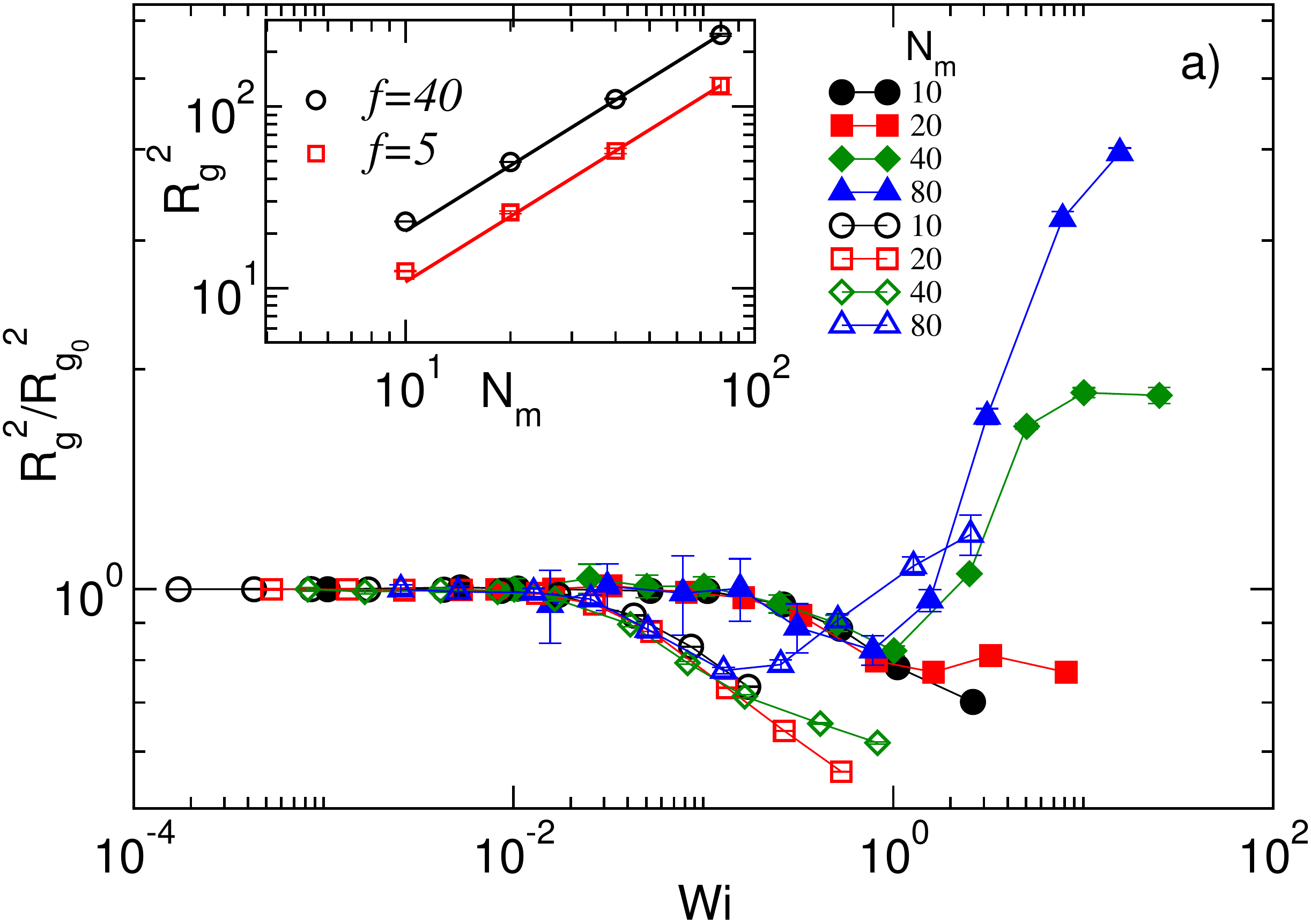}
\includegraphics*[width=\columnwidth]{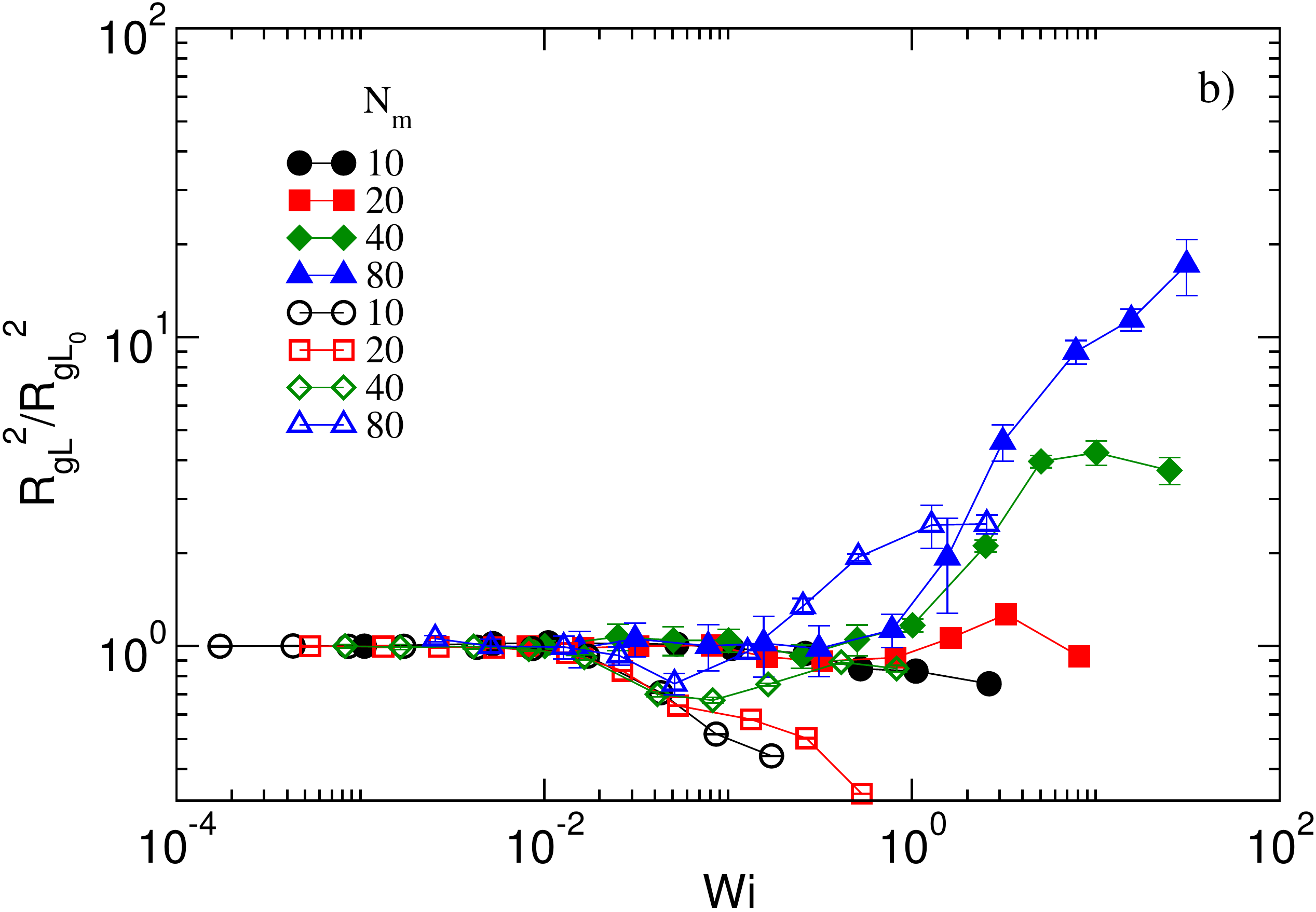}
\includegraphics*[width=\columnwidth]{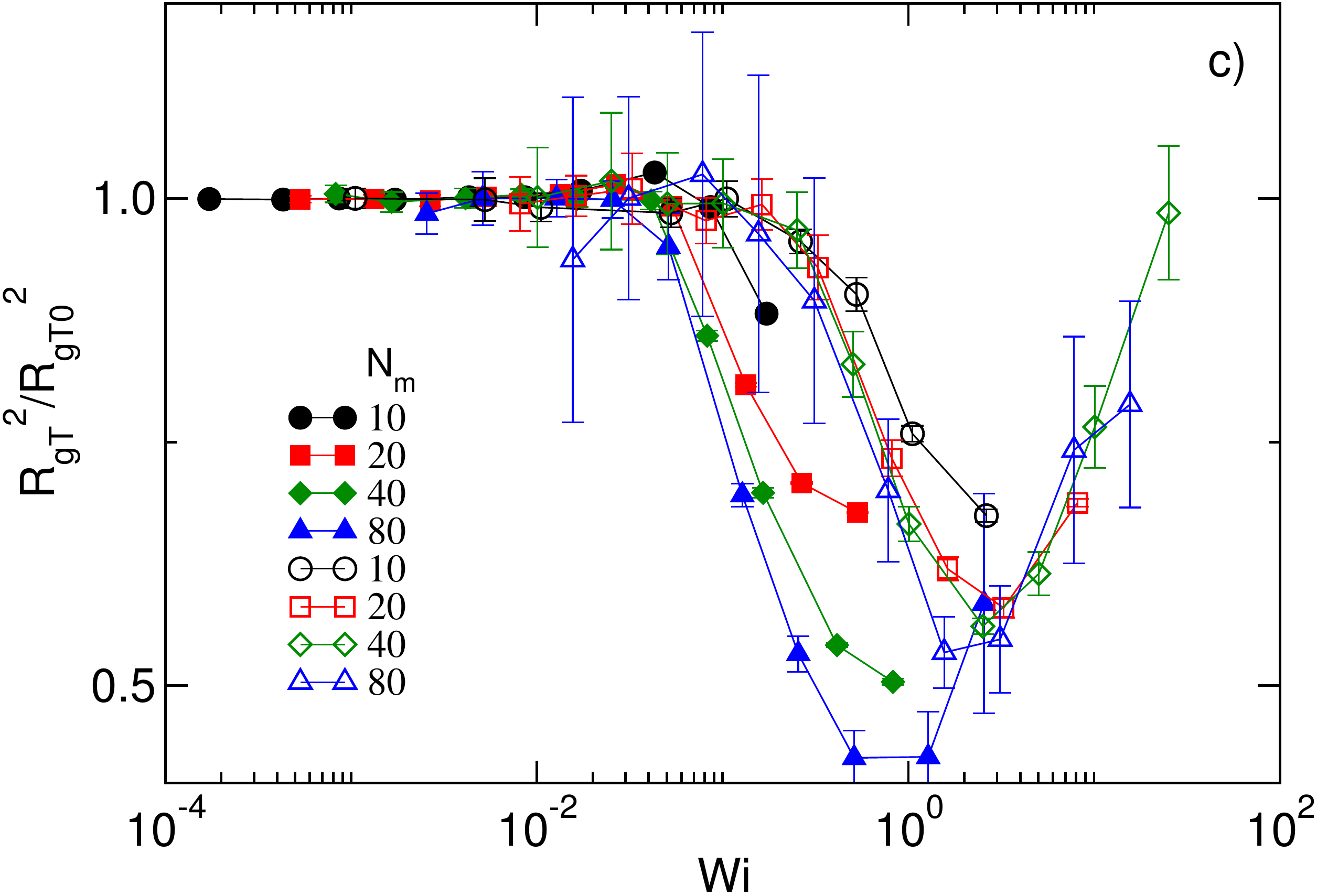}
\caption{a) Overall radii of gyration of star polymers and their components b) along and c) perpendicular to the field as function of $Wi$  for the indicated arm lengths. Open symbols correspond to  $f=40$ and closed symbols to $f=5$. The  inset in a) shows the dependence of the equilibrium radius of gyration $R_{g0}^2 \sim N_m^{2\nu}$ on the arm length $N_m$, where $\nu=0.63$. }
\label{fig:rg_arm_40_5}
\end{figure}

\begin{figure}[t!]
\includegraphics*[width=\columnwidth]{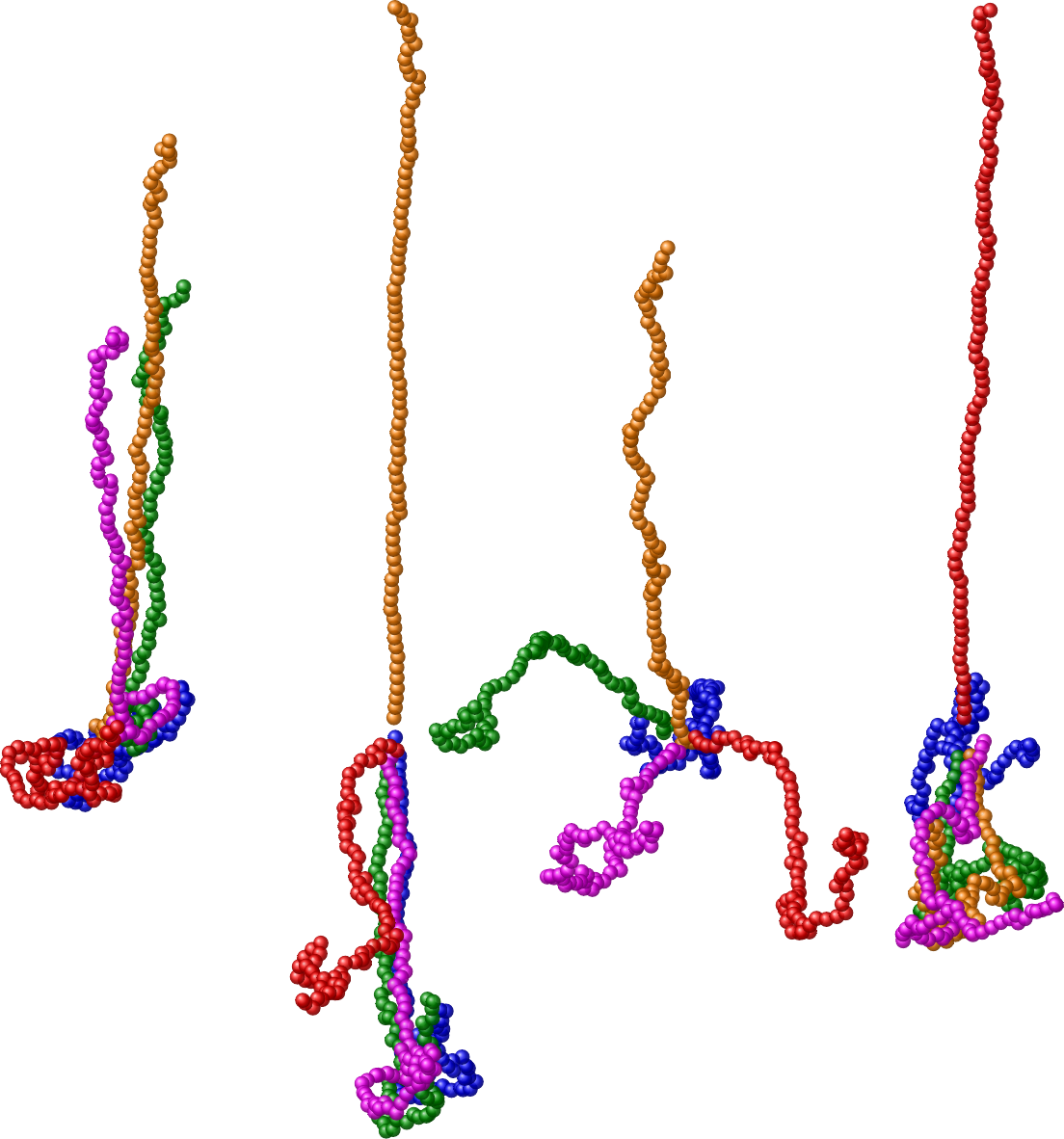}
\includegraphics*[width=\columnwidth]{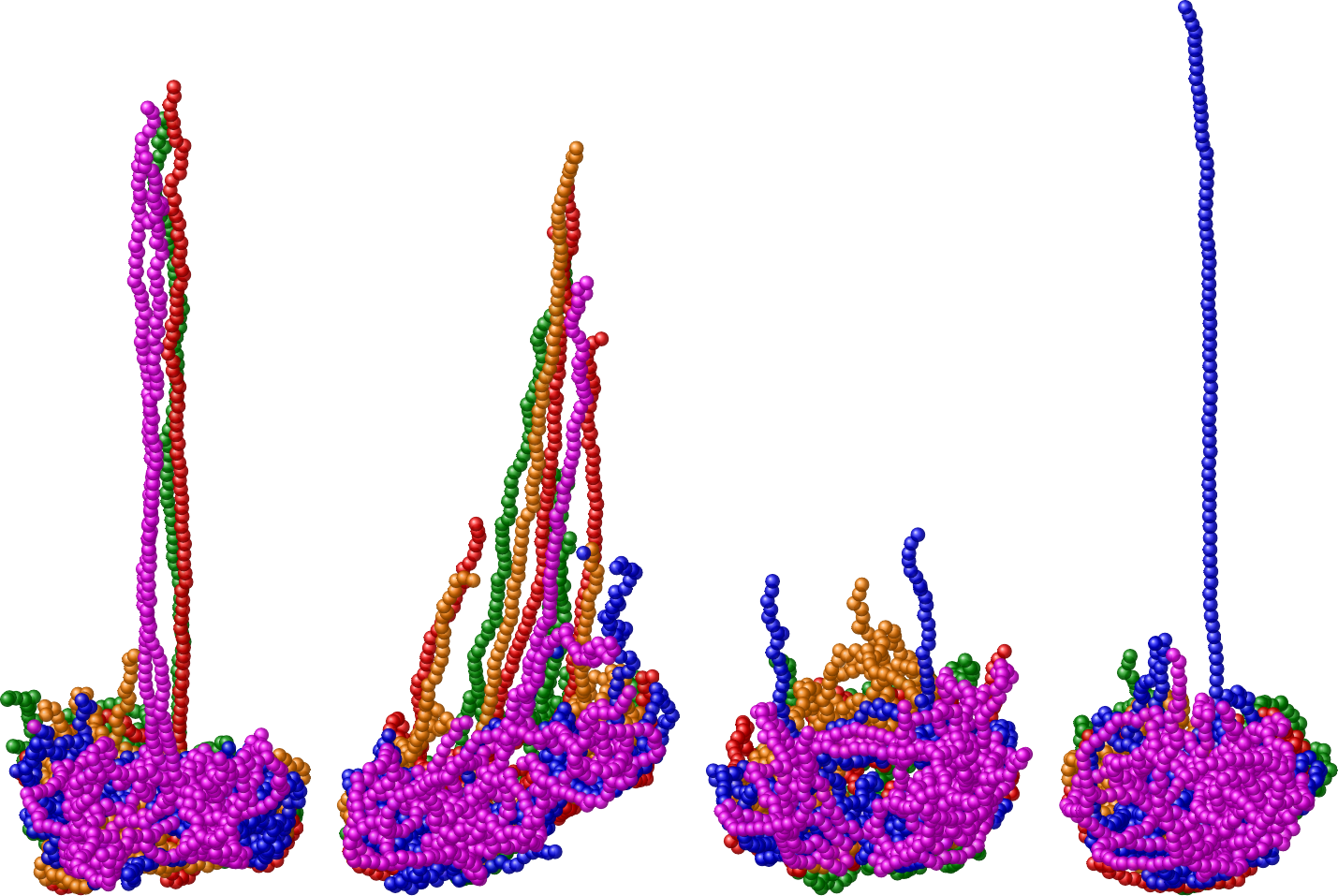}
\caption{Snapshots of sedimenting star polymers for different times. The arm length is $N_m=80$, the strength of the external field  $G=0.5$, and the arm number $f=5$ (top) and $f=40$ (bottom), respectively. See also movies in supporting information.}
\label{fig:snapshot_time}
\end{figure}

Figure~\ref{fig:rg} shows normalized mean square radii of gyration $R_g^2$ and its components $R_{gL}^2$ and $R_{gT}^2$  along and transverse to the external field, respectively, as function of the field strength $G$ for various functionalities. In the linear response regime, $G \lesssim 2 \times 10^{-4}$, the size of a colloid is equal to its unperturb equilibrium value. In an intermediate regime, $R_g^2$ decreases with increasing field strength. This compression of the ultra-soft colloid is more pronounced for high functionality star polymers. Both, the transverse and longitudinal components $R_{gL}^2$ and $R_{gT}^2$ are reduced. In the case of $f=10$, mainly $R_{gT}^2$ decreases with increasing $G$. We attribute this reduction in colloid size to hydrodynamics. The front of the star polymer experience a drag force, which causes a compression. In addition, the flow surrounding the non-draining colloid exerts an inward force, which strongly affects $R_{gT}$. This is similar to the flow field of a linear polymer as discussed in Ref.~\cite{schl:08}. Above a functionality-dependent-field strength, the longitudinal component of the radius of gyration along with $R_{g}$ increases with increasing $G$. This increase is most pronounced for low-functionality star polymers. As illustrated in Fig.~\ref{fig:snap_shot}, in the limit of high fields, polymer arms are stretched, lag behind the center of the star polymer and yield an increase of its radius of gyration. In this regime, the top-bottom symmetry of the star polymer is broken. Polymer arms in front of the star-polymer center are compressed, whereas arms behind the center are stretched significantly. This implies that the monomer density in the front core is higher and therefore also the gravitational pull. A similar anisotropic shape appears for other soft colloidal objects in a gravitational field, such as red blood cells \cite{pelt:13}.
Within the accuracy of our simulations, the components of the radius of gyration tensor seem to approach constant values at large field strengths. (Note the pronounced fluctuations (error bars) of $R^2_{gT}$ at larger $G$ values due to large-scale shape changes of the head (cf. Fig.~\ref{fig:snapshot_time}).)
We attribute this, on the one hand, to the maximal possible stretching of the polymer arms and, on the other hand, a saturation of the compression of the major part of the star polymer by the fluid flow. The latter is to be expected for the transverse component of the radius of gyration, since excluded-volume interactions allow for a minimal size only.

The inset of Fig.~\ref{fig:rg} a) shows the dependence of the equilibrium radius of gyration on the functionality for the arm length $N_m=80$. The solid line indicates the power-law dependence $R_g^2 \sim  f^{1-\nu}$,   with the exponent $\nu \sim 0.63$, which is consistent with the theoretical expectation according to Eq.~(\ref{eq:gyration_radius}).

The dependence of the star polymer radius of gyration on the polymer arm length is displayed in  Fig.~\ref{fig:rg_arm_40_5} for $f=5$ and $f=40$ as function of the Weissenberg number. Here, we find good agreement between the curves for the various arm lengths as long as flow leads to a shrinkage of the star polymers. The appearance of strongly extended tails breaks the universality. Again, the predicted scaling relation (\ref{eq:weissenberg}) fails to describe the obtained functionality dependence.  For short arm lengths, we observe a monotonic decrease of the star polymer size and a crossover to a non-monotonic  behavior for longer arms. In case of short polymers ($N_m \lesssim 20)$, the radius of gyration and its components $R_{gL}$ and  $R_{gT}$ always decreases for all $Wi$ and both functionalities over the considered range of external field strengths. The size of the longer-arm star polymers increases again at higher field strengths due to the appearance of strongly stretched polymers. We expect such an increase for all polymer lengths. There is a critical field strength, which has  to be exceeded to achieve the increase in size.  This critical field strengths depends on the arm length and the functionality and seems to be different for the longitudinal and transverse part of the radius of gyration. Whereas $R_{gL}^2$ clearly increases for $N_m \gtrsim 40$ and $Wi> 1$ ($f=5$) ($Wi>10^{-1}$, ($f=40$)), the respective values for the other arm lengths still decrease. The transverse components of the radii of gyration behave rather similarly, and $R_{gT}^2$ increases again for $Wi \gtrsim 1$. Thereby, the relative change in $R^2_g$ for the fewer-arm stars is always larger than that of the higher functionality stars.

\begin{figure}
\includegraphics*[width=.5\textwidth]{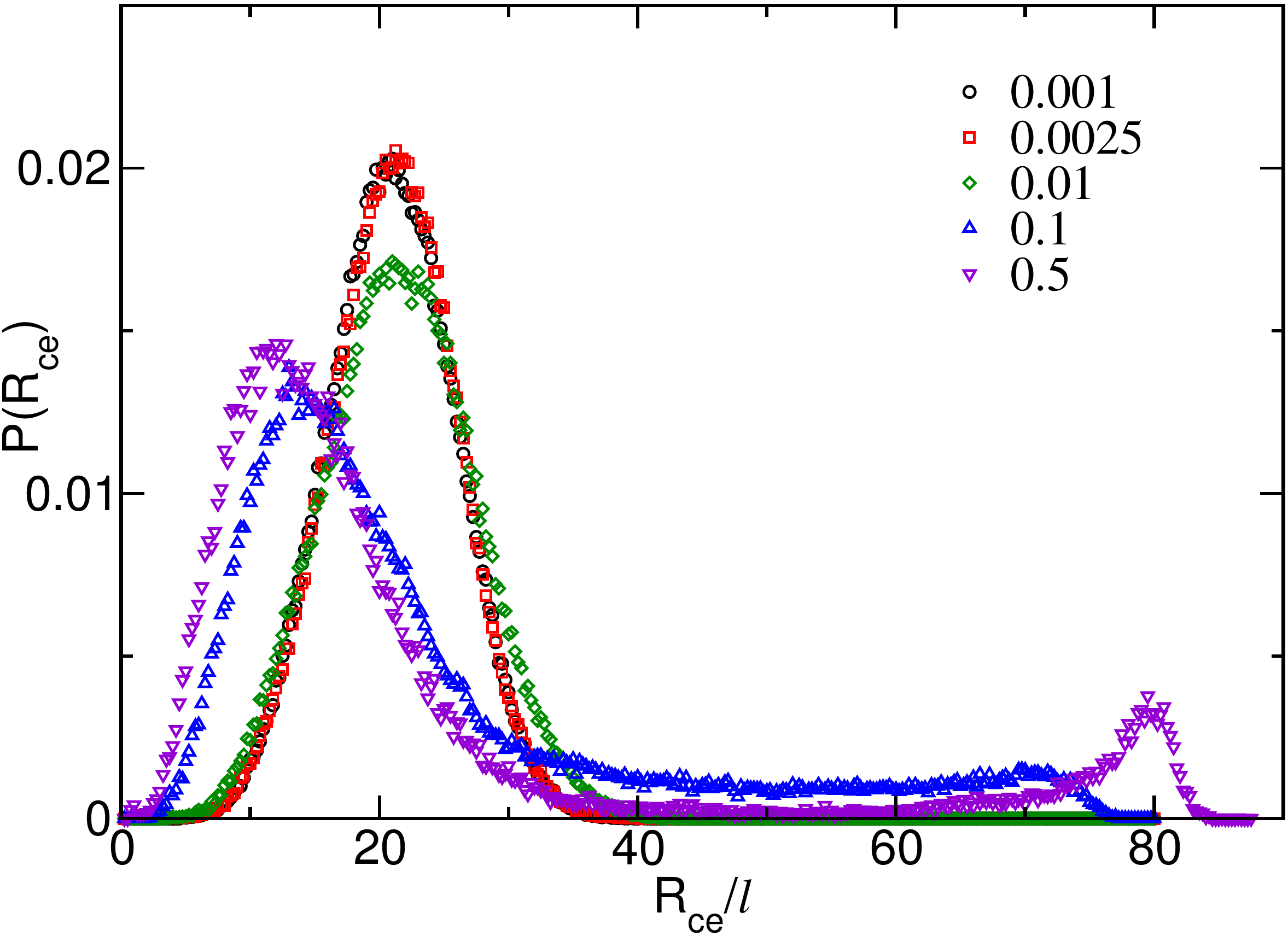}
\caption{Normalized probability distribution function $P(R_{ce})$ of the center-to-end
distance of polymers arms of length $N_m=80$ and the star-polymer functionality $f=40$ for various field strength $G$ (as indicated).}
\label{fig:Pce_40}
\end{figure}

The star-polymer structure is highly dynamic. Although the overall shape is rather stable for a long time  with the majority of polymer arms close to the center of mass  and an extended trailing tail \cite{schl:07,schl:08}, the individual polymers undergo considerable conformational changes.  This is illustrated in  Figure~\ref{fig:snapshot_time}. For a movie, see in supporting information. The emergence of an extended tail leads to an increase of the radii of gyration $R_g^2$ and $R_{gL}^2$. This increase is more pronounced for low-functionality star polymers as reflected in Figs.~\ref{fig:rg} and \ref{fig:rg_arm_40_5}.  For high-functionality star polymers, the relative weight of the small number of arms ($\lesssim 5$) in the trailing tail is less important than for low-functionality star polymers. More remarkable is the increase of $R_{gT}$ at large field strengths. Here, the flow field seems to perturb the lower-field flow-induced compact structure and implies larger conformational changes of the polymers.

We have shown that  the symmetry of the colloidal structure of the star polymer is broken in the limit of high field strengths. The asymmetric distribution of polymer arms in the high-field limit can be qualitatively illustrated by the probability distribution of the center-to-end distance of the polymer arms. As an example, Fig.~\ref{fig:Pce_40} shows the  normalized probability distribution function $P(R_{ce})$ of the center-to-end distance of polymer arms for various field strengths. In the weak-field limit, the distribution of the center-to-end distance exhibits a peak at $R_{ce}/l \approx 20$ corresponding to the equilibrium value.  With increasing  $G$,  the peaks shifts toward  smaller values of the center-to-end distance and broaden substantially for large values of $G$ due to compressive force. At the same time, the probability for extended polymer arms increases. This signifies that on average a few arms are stretched while the majority of arms is compressed.  In the limit of strong fields, the distribution function exhibits two peaks, at $R_{ce}/l \approx 10$ and $R_{ce}/l \approx 80$.
The smaller peak at $R_{ce}/l \approx 80$ corresponds to nearly fully stretched arms. The height of the peak,
smaller than the small-distance peak, reflects that only a few arms are strongly stretched, but that this stretched conformations are rather stable and persistent. Due to fluctuations, stretched arms collapse and are replaced by others. Thereby, the collapse process is very fast, which leads to a low probability in the range $40 \lesssim R_{ce}/l < 70$.

\section{Summary and Conclusions}\label{sec:sec4}

We have investigated the steady-state sedimentation properties of ultra-soft colloids (star polymers) via hybrid mesoscale computer simulations.  We find that the mobility of the ultra-soft colloid exhibits a non-monotonic dependence on the external field strength
$\hat G$. The sedimentation coefficient reaches a maximum value at an intermediate range of $\hat G$. Thereby, for star polymers with fewer
arms, the maximum appears at smaller field strengths and the  sedimentation coefficient assumes smaller values in the high-field regime than the asymptotic value in the limit of vanishing field. The  non-monotonic behavior of the sedimentation coefficient is related to flow-induced conformational changes of the star polymers.  The increase of the sedimentation coefficient follows by a decreases of the radius of gyration.  In the limit of high field strengths, the mobility decreases due to the stretching of various polymer arms along the field direction. Hence, the nonequilibrium dynamical properties of the star polymers are tightly linked with their structure. In the limit of strong external fields, the star polymers are no-longer spherically symmetric. They rather exhibit a compact advancing structure, which is followed by  a trailing tail of a few polymer arms. The number of polymers in the tail strongly fluctuate and their numbers depends on the functionality.

The nonmonotonic behavior of the dynamical and structural properties  appears in our simulations for $Re \gtrsim 1$. Nevertheless, we expect a very similar behavior for Reynolds numbers significantly smaller than unity. Our expectation is supported by the zero-Reynolds number simulations of Refs.~\cite{schl:08,schl:07}, where linear and ring polymers exhibits a qualitatively similar behavior.

For weak fields, the star polymers sediment maintaining their equilibrium shape and the sedimentation coefficient exhibits the arm length and functionality dependence $S_0 \sim N_m^{1-\nu} f^{\delta}$, with $\delta \approx 0.4$. An additional speed up with increasing arm number is obtained for stars in the non-linear regime over a certain range of gravitational constants, with trailing polymer arms. An additional factor is also obtained in the non-linear regime for different arm lengths. However, the increase or even decrease of $S$ is less pronounced by varying the arm length than by varying the functionality. Hence, non-linear effects can enhance sedimentation and promote separation of star polymers of different sizes.

\section{acknowledgement}
 Financial support by the Deutsche Forschungsgemeinschaft (DFG) through the Collaborative Research Center "Physics of Colloidal Dispersions in External Fields" (SFB TR6), by the EU through the Collaborative Research Project "NanoDirect" (NMP4-SL-2008-213948), and the EU through FP7-Infrastructure ESMI (Grant 262348) are gratefully acknowledged. The authors gratefully acknowledge the computing time granted on the supercomputers at J\"{u}lich Supercomputing Centre (JSC).

\appendix

\section{Fluid backflow}

The equation of motion of a monomer of the star polymer in the laboratory reference frame (indicated by a prime) is given by
\begin{align} \label{eqa:eom}
M {\ddot{\bm R}}_k^{'\mu} = \bm F_k^{\mu} + M \hat{\bm G} ,
\end{align}
where the $\bm F_k^{\mu}$ are intramolecular forces following from the potential (\ref{eq:pot_bond}) and (\ref{eq:pot_int}). The center-of-mass velocity of the total system, star polymer plus MPC fluid, is then
\begin{align} \label{eqa:com}
M_{tot} \ddot{\bm r}'_{cm} = N_p M \hat{\bm G} ,
\end{align}
with $M_{tot}=N_p M + N_s m$ and
\begin{align}
\bm r'_{cm} = \frac{1}{M_{tot}} \left(\sum_{k, \mu} M \bm R_k^{'\mu} + \sum_i m \bm r'_i  \right) .
\end{align}
Introducing the coordinates ${\bm R}_k^{\mu} = \bm R_k^{'\mu} - \bm r'_{cm}$ and ${\bm r}_i = \bm r'_i - \bm r'_{cm}$
of the monomer and fluid particle positions with respect to the total center-of-mass implies
\begin{align}
\sum_{k, \mu} M \ddot{\bm R}_k^{\mu} + \sum_i m \ddot{\bm r}_i  =  0 ,
\end{align}
i.e., conservation of the total moment in the center-of-mass reference frame. We set this moment to zero initially.  From Eq.~(\ref{eqa:eom}), we obtain then
\begin{align}
M \ddot{\bm R}_k^{\mu} = \bm F_k^{\mu} + M \hat{\bm G} + \bm F_f ,
\end{align}
with
\begin{align} \label{eqa:force_back_flow}
{\bm F}_f=-\frac{M^2 N_p}{M N_p + mN_s} \hat{\bm G}  .
\end{align}
Similarly, we obtain for the fluid particles
\begin{align}
m \ddot{\bm r}_i = \frac{m }{M }  {\bm F}_f .
\end{align}

\section*{Supplementary Material}\label{sec:suppinfo}
Two movie files and a figure are provided to illustrate the star polymer conformations.
 S1: A sedimenting star polymer is shown w.r.t. to its centre bead for the functionality $f=5$ and $G=0.5$ at the arm length $N_m=80$. S2: The same set of parameters are applied as in $S1$, except the  number of arms is $f=40$. \\



\begin{thebibliography}{92}
\expandafter\ifx\csname natexlab\endcsname\relax\def\natexlab#1{#1}\fi
\expandafter\ifx\csname bibnamefont\endcsname\relax
  \def\bibnamefont#1{#1}\fi
\expandafter\ifx\csname bibfnamefont\endcsname\relax
  \def\bibfnamefont#1{#1}\fi
\expandafter\ifx\csname citenamefont\endcsname\relax
  \def\citenamefont#1{#1}\fi
\expandafter\ifx\csname url\endcsname\relax
  \def\url#1{\texttt{#1}}\fi
\expandafter\ifx\csname urlprefix\endcsname\relax\def\urlprefix{URL }\fi
\providecommand{\bibinfo}[2]{#2}
\providecommand{\eprint}[2][]{\url{#2}}

\bibitem[{\citenamefont{Larson}(1999)}]{Larson_SRF_1999}
\bibinfo{author}{\bibfnamefont{R.~G.} \bibnamefont{Larson}},
  \emph{\bibinfo{title}{The structure and rheology of complex fluids}}
  (\bibinfo{publisher}{Oxford University Press}, \bibinfo{address}{Oxford, NY},
  \bibinfo{year}{1999}).

\bibitem[{\citenamefont{Bird et~al.}(1987)\citenamefont{Bird, Armstrong, and
  Hassager}}]{Bird_DPL_1987}
\bibinfo{author}{\bibfnamefont{R.~B.} \bibnamefont{Bird}},
  \bibinfo{author}{\bibfnamefont{R.~C.} \bibnamefont{Armstrong}},
  \bibnamefont{and} \bibinfo{author}{\bibfnamefont{O.}~\bibnamefont{Hassager}},
  \emph{\bibinfo{title}{Dynamics of polymeric liquids:{F}luid mechanics}}
  (\bibinfo{publisher}{Wiley}, \bibinfo{address}{New York},
  \bibinfo{year}{1987}), \bibinfo{edition}{2nd} ed.

\bibitem[{\citenamefont{Smith et~al.}(1999)\citenamefont{Smith, Babcock, and
  Chu}}]{smit:99}
\bibinfo{author}{\bibfnamefont{D.~E.} \bibnamefont{Smith}},
  \bibinfo{author}{\bibfnamefont{H.~P.} \bibnamefont{Babcock}},
  \bibnamefont{and} \bibinfo{author}{\bibfnamefont{S.}~\bibnamefont{Chu}},
  \bibinfo{journal}{Science} \textbf{\bibinfo{volume}{283}},
  \bibinfo{pages}{1724} (\bibinfo{year}{1999}).

\bibitem[{\citenamefont{LeDuc et~al.}(1999)\citenamefont{LeDuc, Haber, Boa, and
  Wirtz}}]{ledu:99}
\bibinfo{author}{\bibfnamefont{P.}~\bibnamefont{LeDuc}},
  \bibinfo{author}{\bibfnamefont{C.}~\bibnamefont{Haber}},
  \bibinfo{author}{\bibfnamefont{G.}~\bibnamefont{Boa}}, \bibnamefont{and}
  \bibinfo{author}{\bibfnamefont{D.}~\bibnamefont{Wirtz}},
  \bibinfo{journal}{Nature} \textbf{\bibinfo{volume}{399}},
  \bibinfo{pages}{564} (\bibinfo{year}{1999}).

\bibitem[{\citenamefont{Kr{\"o}ger}(2004)}]{kroe:04}
\bibinfo{author}{\bibfnamefont{M.}~\bibnamefont{Kr{\"o}ger}},
  \bibinfo{journal}{Phys. Rep.} \textbf{\bibinfo{volume}{390}},
  \bibinfo{pages}{453} (\bibinfo{year}{2004}).

\bibitem[{\citenamefont{Schroeder et~al.}(2005)\citenamefont{Schroeder,
  Teixeira, Shaqfeh, and Chu}}]{schr:05}
\bibinfo{author}{\bibfnamefont{C.~M.} \bibnamefont{Schroeder}},
  \bibinfo{author}{\bibfnamefont{R.~E.} \bibnamefont{Teixeira}},
  \bibinfo{author}{\bibfnamefont{E.~S.~G.} \bibnamefont{Shaqfeh}},
  \bibnamefont{and} \bibinfo{author}{\bibfnamefont{S.}~\bibnamefont{Chu}},
  \bibinfo{journal}{Phys. Rev. Lett.} \textbf{\bibinfo{volume}{95}},
  \bibinfo{pages}{018301} (\bibinfo{year}{2005}).

\bibitem[{\citenamefont{Gerashchenko and Steinberg}(2006)}]{gera:06}
\bibinfo{author}{\bibfnamefont{S.}~\bibnamefont{Gerashchenko}}
  \bibnamefont{and}
  \bibinfo{author}{\bibfnamefont{V.}~\bibnamefont{Steinberg}},
  \bibinfo{journal}{Phys. Rev. Lett.} \textbf{\bibinfo{volume}{96}},
  \bibinfo{pages}{038304} (\bibinfo{year}{2006}).

\bibitem[{\citenamefont{Winkler}(2006)}]{wink:06.1}
\bibinfo{author}{\bibfnamefont{R.~G.} \bibnamefont{Winkler}},
  \bibinfo{journal}{Phys. Rev. Lett.} \textbf{\bibinfo{volume}{97}},
  \bibinfo{pages}{128301} (\bibinfo{year}{2006}).

\bibitem[{\citenamefont{Jendrejack et~al.}(2004)\citenamefont{Jendrejack,
  Schwartz, de~Pablo, and Graham}}]{jend:04}
\bibinfo{author}{\bibfnamefont{R.~M.} \bibnamefont{Jendrejack}},
  \bibinfo{author}{\bibfnamefont{D.~C.} \bibnamefont{Schwartz}},
  \bibinfo{author}{\bibfnamefont{J.~J.} \bibnamefont{de~Pablo}},
  \bibnamefont{and} \bibinfo{author}{\bibfnamefont{M.~D.}
  \bibnamefont{Graham}}, \bibinfo{journal}{J. Chem. Phys.}
  \textbf{\bibinfo{volume}{120}}, \bibinfo{pages}{2513} (\bibinfo{year}{2004}).

\bibitem[{\citenamefont{Chelakkot et~al.}(2010)\citenamefont{Chelakkot,
  Winkler, and Gompper}}]{chel:10}
\bibinfo{author}{\bibfnamefont{R.}~\bibnamefont{Chelakkot}},
  \bibinfo{author}{\bibfnamefont{R.~G.} \bibnamefont{Winkler}},
  \bibnamefont{and} \bibinfo{author}{\bibfnamefont{G.}~\bibnamefont{Gompper}},
  \bibinfo{journal}{EPL} \textbf{\bibinfo{volume}{91}}, \bibinfo{pages}{14001}
  (\bibinfo{year}{2010}).

\bibitem[{\citenamefont{Chelakkot et~al.}(2012)\citenamefont{Chelakkot,
  Winkler, and Gompper}}]{chel:12}
\bibinfo{author}{\bibfnamefont{R.}~\bibnamefont{Chelakkot}},
  \bibinfo{author}{\bibfnamefont{R.~G.} \bibnamefont{Winkler}},
  \bibnamefont{and} \bibinfo{author}{\bibfnamefont{G.}~\bibnamefont{Gompper}},
  \bibinfo{journal}{Phys. Rev. Lett.} \textbf{\bibinfo{volume}{109}},
  \bibinfo{pages}{178101} (\bibinfo{year}{2012}).

\bibitem[{\citenamefont{Steinhauser et~al.}(2012)\citenamefont{Steinhauser,
  K{\"o}ster, and Pfohl}}]{stei:12}
\bibinfo{author}{\bibfnamefont{D.}~\bibnamefont{Steinhauser}},
  \bibinfo{author}{\bibfnamefont{S.}~\bibnamefont{K{\"o}ster}},
  \bibnamefont{and} \bibinfo{author}{\bibfnamefont{T.}~\bibnamefont{Pfohl}},
  \bibinfo{journal}{ACS Macro Lett.} \textbf{\bibinfo{volume}{1}},
  \bibinfo{pages}{541} (\bibinfo{year}{2012}).

\bibitem[{\citenamefont{Harasim et~al.}(2013)\citenamefont{Harasim, Wunderlich,
  Peleg, Kr\"oger, and Bausch}}]{hara:13}
\bibinfo{author}{\bibfnamefont{M.}~\bibnamefont{Harasim}},
  \bibinfo{author}{\bibfnamefont{B.}~\bibnamefont{Wunderlich}},
  \bibinfo{author}{\bibfnamefont{O.}~\bibnamefont{Peleg}},
  \bibinfo{author}{\bibfnamefont{M.}~\bibnamefont{Kr\"oger}}, \bibnamefont{and}
  \bibinfo{author}{\bibfnamefont{A.~R.} \bibnamefont{Bausch}},
  \bibinfo{journal}{Phys. Rev. Lett.} \textbf{\bibinfo{volume}{110}},
  \bibinfo{pages}{108302} (\bibinfo{year}{2013}).

\bibitem[{\citenamefont{Pakula et~al.}(1998)\citenamefont{Pakula, Vlassopoulos,
  Fytas, and Roovers}}]{paku:98}
\bibinfo{author}{\bibfnamefont{T.}~\bibnamefont{Pakula}},
  \bibinfo{author}{\bibfnamefont{D.}~\bibnamefont{Vlassopoulos}},
  \bibinfo{author}{\bibfnamefont{G.}~\bibnamefont{Fytas}}, \bibnamefont{and}
  \bibinfo{author}{\bibfnamefont{J.}~\bibnamefont{Roovers}},
  \bibinfo{journal}{Macromolecules} \textbf{\bibinfo{volume}{31}},
  \bibinfo{pages}{8931} (\bibinfo{year}{1998}).

\bibitem[{\citenamefont{Vlassopoulos et~al.}(2001)\citenamefont{Vlassopoulos,
  Fytas, Pakula, and Roovers}}]{Vlassopoulos_MAS_2001}
\bibinfo{author}{\bibfnamefont{D.}~\bibnamefont{Vlassopoulos}},
  \bibinfo{author}{\bibfnamefont{G.}~\bibnamefont{Fytas}},
  \bibinfo{author}{\bibfnamefont{T.}~\bibnamefont{Pakula}}, \bibnamefont{and}
  \bibinfo{author}{\bibfnamefont{J.}~\bibnamefont{Roovers}},
  \bibinfo{journal}{J. Phys. Condens. Matter} \textbf{\bibinfo{volume}{13}},
  \bibinfo{pages}{R855} (\bibinfo{year}{2001}).

\bibitem[{\citenamefont{Nikoubashman and
  Likos}(2010{\natexlab{a}})}]{niko:10.1}
\bibinfo{author}{\bibfnamefont{A.}~\bibnamefont{Nikoubashman}}
  \bibnamefont{and} \bibinfo{author}{\bibfnamefont{C.~N.} \bibnamefont{Likos}},
  \bibinfo{journal}{Macromolecules} \textbf{\bibinfo{volume}{43}},
  \bibinfo{pages}{1610} (\bibinfo{year}{2010}{\natexlab{a}}),
  \urlprefix\url{http://dx.doi.org/10.1021/ma902212s}.

\bibitem[{\citenamefont{Vlassopoulos and Cloitre}(2014)}]{vlas:14}
\bibinfo{author}{\bibfnamefont{D.}~\bibnamefont{Vlassopoulos}}
  \bibnamefont{and} \bibinfo{author}{\bibfnamefont{M.}~\bibnamefont{Cloitre}},
  \bibinfo{journal}{Curr. Opin. Colloid Interface Sci.}
  \textbf{\bibinfo{volume}{19}}, \bibinfo{pages}{561} (\bibinfo{year}{2014}).

\bibitem[{\citenamefont{Winkler et~al.}(2014)\citenamefont{Winkler, Fedosov,
  and Gompper}}]{wink:14.1}
\bibinfo{author}{\bibfnamefont{R.~G.} \bibnamefont{Winkler}},
  \bibinfo{author}{\bibfnamefont{D.~A.} \bibnamefont{Fedosov}},
  \bibnamefont{and} \bibinfo{author}{\bibfnamefont{G.}~\bibnamefont{Gompper}},
  \bibinfo{journal}{Curr. Opin. Colloid Interface Sci.}
  \textbf{\bibinfo{volume}{19}}, \bibinfo{pages}{594} (\bibinfo{year}{2014}),
  \urlprefix\url{http://www.sciencedirect.com/science/article/pii/S1359029414000922}.

\bibitem[{\citenamefont{Keller and Skalak}(1982)}]{kell82}
\bibinfo{author}{\bibfnamefont{S.~R.} \bibnamefont{Keller}} \bibnamefont{and}
  \bibinfo{author}{\bibfnamefont{R.}~\bibnamefont{Skalak}},
  \bibinfo{journal}{Journal of Fluid Mechanics} \textbf{\bibinfo{volume}{120}},
  \bibinfo{pages}{27} (\bibinfo{year}{1982}).

\bibitem[{\citenamefont{Noguchi and Gompper}(2004)}]{nogu:04}
\bibinfo{author}{\bibfnamefont{H.}~\bibnamefont{Noguchi}} \bibnamefont{and}
  \bibinfo{author}{\bibfnamefont{G.}~\bibnamefont{Gompper}},
  \bibinfo{journal}{Phys. Rev. Lett.} \textbf{\bibinfo{volume}{93}},
  \bibinfo{pages}{258102} (\bibinfo{year}{2004}).

\bibitem[{\citenamefont{Kantsler and Steinberg}(2006)}]{kant06}
\bibinfo{author}{\bibfnamefont{V.}~\bibnamefont{Kantsler}} \bibnamefont{and}
  \bibinfo{author}{\bibfnamefont{V.}~\bibnamefont{Steinberg}},
  \bibinfo{journal}{Physical review letters} \textbf{\bibinfo{volume}{96}},
  \bibinfo{pages}{036001} (\bibinfo{year}{2006}).

\bibitem[{\citenamefont{Misbah}(2006)}]{misb06}
\bibinfo{author}{\bibfnamefont{C.}~\bibnamefont{Misbah}},
  \bibinfo{journal}{Physical review letters} \textbf{\bibinfo{volume}{96}},
  \bibinfo{pages}{028104} (\bibinfo{year}{2006}).

\bibitem[{\citenamefont{Lebedev et~al.}(2007)\citenamefont{Lebedev, Turitsyn,
  and Vergeles}}]{lebe07}
\bibinfo{author}{\bibfnamefont{V.}~\bibnamefont{Lebedev}},
  \bibinfo{author}{\bibfnamefont{K.}~\bibnamefont{Turitsyn}}, \bibnamefont{and}
  \bibinfo{author}{\bibfnamefont{S.}~\bibnamefont{Vergeles}},
  \bibinfo{journal}{Physical review letters} \textbf{\bibinfo{volume}{99}},
  \bibinfo{pages}{218101} (\bibinfo{year}{2007}).

\bibitem[{\citenamefont{Vlahovska and Gracia}(2007)}]{vlah07}
\bibinfo{author}{\bibfnamefont{P.~M.} \bibnamefont{Vlahovska}}
  \bibnamefont{and} \bibinfo{author}{\bibfnamefont{R.~S.}
  \bibnamefont{Gracia}}, \bibinfo{journal}{Physical Review E}
  \textbf{\bibinfo{volume}{75}}, \bibinfo{pages}{016313}
  (\bibinfo{year}{2007}).

\bibitem[{\citenamefont{Zhao and Shaqfeh}(2011)}]{zhao11}
\bibinfo{author}{\bibfnamefont{H.}~\bibnamefont{Zhao}} \bibnamefont{and}
  \bibinfo{author}{\bibfnamefont{E.~S.} \bibnamefont{Shaqfeh}},
  \bibinfo{journal}{Journal of Fluid Mechanics} \textbf{\bibinfo{volume}{674}},
  \bibinfo{pages}{578} (\bibinfo{year}{2011}).

\bibitem[{\citenamefont{Abreu et~al.}(2014)\citenamefont{Abreu, Levant,
  Steinberg, and Seifert}}]{abre14}
\bibinfo{author}{\bibfnamefont{D.}~\bibnamefont{Abreu}},
  \bibinfo{author}{\bibfnamefont{M.}~\bibnamefont{Levant}},
  \bibinfo{author}{\bibfnamefont{V.}~\bibnamefont{Steinberg}},
  \bibnamefont{and} \bibinfo{author}{\bibfnamefont{U.}~\bibnamefont{Seifert}},
  \bibinfo{journal}{Advances in colloid and interface science}
  \textbf{\bibinfo{volume}{208}}, \bibinfo{pages}{129} (\bibinfo{year}{2014}).

\bibitem[{\citenamefont{Lamura and Gompper}(2013)}]{lamu13}
\bibinfo{author}{\bibfnamefont{A.}~\bibnamefont{Lamura}} \bibnamefont{and}
  \bibinfo{author}{\bibfnamefont{G.}~\bibnamefont{Gompper}},
  \bibinfo{journal}{EPL (Europhysics Letters)} \textbf{\bibinfo{volume}{102}},
  \bibinfo{pages}{28004} (\bibinfo{year}{2013}).

\bibitem[{\citenamefont{Abkarian et~al.}(2007)\citenamefont{Abkarian, Faivre,
  and Viallat}}]{Abkarian_SSF_2007}
\bibinfo{author}{\bibfnamefont{M.}~\bibnamefont{Abkarian}},
  \bibinfo{author}{\bibfnamefont{M.}~\bibnamefont{Faivre}}, \bibnamefont{and}
  \bibinfo{author}{\bibfnamefont{A.}~\bibnamefont{Viallat}},
  \bibinfo{journal}{Physical Review Letters} \textbf{\bibinfo{volume}{98}},
  \bibinfo{pages}{188302} (\bibinfo{year}{2007}).

\bibitem[{\citenamefont{Noguchi and Gompper}(2005)}]{nogu:05}
\bibinfo{author}{\bibfnamefont{H.}~\bibnamefont{Noguchi}} \bibnamefont{and}
  \bibinfo{author}{\bibfnamefont{G.}~\bibnamefont{Gompper}},
  \bibinfo{journal}{Proceedings of the National Academy of Sciences of the
  United States of America} \textbf{\bibinfo{volume}{102}},
  \bibinfo{pages}{14159} (\bibinfo{year}{2005}).

\bibitem[{\citenamefont{Kaoui et~al.}(2009)\citenamefont{Kaoui, Biros, and
  Misbah}}]{kaou:09}
\bibinfo{author}{\bibfnamefont{B.}~\bibnamefont{Kaoui}},
  \bibinfo{author}{\bibfnamefont{G.}~\bibnamefont{Biros}}, \bibnamefont{and}
  \bibinfo{author}{\bibfnamefont{C.}~\bibnamefont{Misbah}},
  \bibinfo{journal}{Physical review letters} \textbf{\bibinfo{volume}{103}},
  \bibinfo{pages}{188101} (\bibinfo{year}{2009}).

\bibitem[{\citenamefont{Dupire et~al.}(2012)\citenamefont{Dupire, Socol, and
  Viallat}}]{Dupire_DRC_2012}
\bibinfo{author}{\bibfnamefont{J.}~\bibnamefont{Dupire}},
  \bibinfo{author}{\bibfnamefont{M.}~\bibnamefont{Socol}}, \bibnamefont{and}
  \bibinfo{author}{\bibfnamefont{A.}~\bibnamefont{Viallat}},
  \bibinfo{journal}{Proceedings of the National Academy of Sciences}
  \textbf{\bibinfo{volume}{109}}, \bibinfo{pages}{20808}
  (\bibinfo{year}{2012}).

\bibitem[{\citenamefont{Pivkin and Karniadakis}(2008)}]{Pivkin_ACG_2008}
\bibinfo{author}{\bibfnamefont{I.~V.} \bibnamefont{Pivkin}} \bibnamefont{and}
  \bibinfo{author}{\bibfnamefont{G.~E.} \bibnamefont{Karniadakis}},
  \bibinfo{journal}{Physical Review Letters} \textbf{\bibinfo{volume}{101}},
  \bibinfo{pages}{118105} (\bibinfo{year}{2008}).

\bibitem[{\citenamefont{McWhirter et~al.}(2009)\citenamefont{McWhirter,
  Noguchi, and Gompper}}]{McWhirter_FIC_2009}
\bibinfo{author}{\bibfnamefont{J.~L.} \bibnamefont{McWhirter}},
  \bibinfo{author}{\bibfnamefont{H.}~\bibnamefont{Noguchi}}, \bibnamefont{and}
  \bibinfo{author}{\bibfnamefont{G.}~\bibnamefont{Gompper}},
  \bibinfo{journal}{Proceedings of the National Academy of Sciences USA}
  \textbf{\bibinfo{volume}{106}}, \bibinfo{pages}{6039} (\bibinfo{year}{2009}).

\bibitem[{\citenamefont{Tomaiuolo et~al.}(2009)\citenamefont{Tomaiuolo,
  Simeone, Martinelli, Rotoli, and Guido}}]{toma09}
\bibinfo{author}{\bibfnamefont{G.}~\bibnamefont{Tomaiuolo}},
  \bibinfo{author}{\bibfnamefont{M.}~\bibnamefont{Simeone}},
  \bibinfo{author}{\bibfnamefont{V.}~\bibnamefont{Martinelli}},
  \bibinfo{author}{\bibfnamefont{B.}~\bibnamefont{Rotoli}}, \bibnamefont{and}
  \bibinfo{author}{\bibfnamefont{S.}~\bibnamefont{Guido}},
  \bibinfo{journal}{Soft Matter} \textbf{\bibinfo{volume}{5}},
  \bibinfo{pages}{3736} (\bibinfo{year}{2009}).

\bibitem[{\citenamefont{Reasor et~al.}(2012)\citenamefont{Reasor, Clausen, and
  Aidun}}]{Reasor_CLB_2012}
\bibinfo{author}{\bibfnamefont{D.~A.} \bibnamefont{Reasor}},
  \bibinfo{author}{\bibfnamefont{J.~R.} \bibnamefont{Clausen}},
  \bibnamefont{and} \bibinfo{author}{\bibfnamefont{C.~K.} \bibnamefont{Aidun}},
  \bibinfo{journal}{International Journal for Numerical Methods in Fluids}
  \textbf{\bibinfo{volume}{68}}, \bibinfo{pages}{767} (\bibinfo{year}{2012}).

\bibitem[{\citenamefont{Fedosov et~al.}(2014)\citenamefont{Fedosov,
  Peltom{\"a}ki, and Gompper}}]{Fedosov_DDC_2014}
\bibinfo{author}{\bibfnamefont{D.~A.} \bibnamefont{Fedosov}},
  \bibinfo{author}{\bibfnamefont{M.}~\bibnamefont{Peltom{\"a}ki}},
  \bibnamefont{and} \bibinfo{author}{\bibfnamefont{G.}~\bibnamefont{Gompper}},
  \bibinfo{journal}{Soft matter} \textbf{\bibinfo{volume}{10}},
  \bibinfo{pages}{4258} (\bibinfo{year}{2014}).

\bibitem[{\citenamefont{Peltomaki and Gompper}(2013)}]{pelt:13}
\bibinfo{author}{\bibfnamefont{M.}~\bibnamefont{Peltomaki}} \bibnamefont{and}
  \bibinfo{author}{\bibfnamefont{G.}~\bibnamefont{Gompper}},
  \bibinfo{journal}{Soft Matter} \textbf{\bibinfo{volume}{9}},
  \bibinfo{pages}{8346} (\bibinfo{year}{2013}),
  \urlprefix\url{http://dx.doi.org/10.1039/C3SM50592H}.

\bibitem[{\citenamefont{Harding et~al.}(1993)\citenamefont{Harding, Rowe, and
  Horton}}]{Harding_PS_1993}
\bibinfo{author}{\bibfnamefont{S.~E.} \bibnamefont{Harding}},
  \bibinfo{author}{\bibfnamefont{A.~J.} \bibnamefont{Rowe}}, \bibnamefont{and}
  \bibinfo{author}{\bibfnamefont{J.~C.} \bibnamefont{Horton}},
  \emph{\bibinfo{title}{Analytical Ultracentrifugation in Biochemistry and
  Polymer Science; Royal Society of Chemistry}} (\bibinfo{publisher}{Cambridge,
  UK}, \bibinfo{year}{1993}).

\bibitem[{\citenamefont{Laue and Stafford}(1993)}]{Laue_ARBBS_1999}
\bibinfo{author}{\bibfnamefont{T.~M.} \bibnamefont{Laue}} \bibnamefont{and}
  \bibinfo{author}{\bibfnamefont{W.~F.} \bibnamefont{Stafford}},
  \bibinfo{journal}{Annu. ReV. Biophys. Biomol. Struct.}
  \textbf{\bibinfo{volume}{28}}, \bibinfo{pages}{75} (\bibinfo{year}{1993}).

\bibitem[{\citenamefont{Abbitt and Nash}(2003)}]{Abbitt_RPB_2003}
\bibinfo{author}{\bibfnamefont{K.~B.} \bibnamefont{Abbitt}} \bibnamefont{and}
  \bibinfo{author}{\bibfnamefont{G.~B.} \bibnamefont{Nash}},
  \bibinfo{journal}{Am. J. Physiology} \textbf{\bibinfo{volume}{285}},
  \bibinfo{pages}{H229} (\bibinfo{year}{2003}).

\bibitem[{\citenamefont{Schlagberger and Netz}(2008)}]{schl:08}
\bibinfo{author}{\bibfnamefont{X.}~\bibnamefont{Schlagberger}}
  \bibnamefont{and} \bibinfo{author}{\bibfnamefont{R.~R.} \bibnamefont{Netz}},
  \bibinfo{journal}{Macromolecules} \textbf{\bibinfo{volume}{41}},
  \bibinfo{pages}{1861} (\bibinfo{year}{2008}),
  \urlprefix\url{http://dx.doi.org/10.1021/ma070947m}.

\bibitem[{\citenamefont{Zimm and Schumaker}(1976)}]{zimm:76}
\bibinfo{author}{\bibfnamefont{B.}~\bibnamefont{Zimm}} \bibnamefont{and}
  \bibinfo{author}{\bibfnamefont{V.}~\bibnamefont{Schumaker}},
  \bibinfo{journal}{Biophys. Chem.} \textbf{\bibinfo{volume}{5}},
  \bibinfo{pages}{265} (\bibinfo{year}{1976}).

\bibitem[{\citenamefont{Erta\ifmmode~\mbox{\c{s}}\else \c{s}\fi{} and
  Kardar}(1993)}]{erta:93}
\bibinfo{author}{\bibfnamefont{D.}~\bibnamefont{Erta\ifmmode~\mbox{\c{s}}\else
  \c{s}\fi{}}} \bibnamefont{and}
  \bibinfo{author}{\bibfnamefont{M.}~\bibnamefont{Kardar}},
  \bibinfo{journal}{Phys. Rev. E} \textbf{\bibinfo{volume}{48}},
  \bibinfo{pages}{1228} (\bibinfo{year}{1993}),
  \urlprefix\url{http://link.aps.org/doi/10.1103/PhysRevE.48.1228}.

\bibitem[{\citenamefont{Schlagberger and Netz}(2007)}]{schl:07}
\bibinfo{author}{\bibfnamefont{X.}~\bibnamefont{Schlagberger}}
  \bibnamefont{and} \bibinfo{author}{\bibfnamefont{R.~R.} \bibnamefont{Netz}},
  \bibinfo{journal}{Phys. Rev. Lett.} \textbf{\bibinfo{volume}{98}},
  \bibinfo{pages}{128301} (\bibinfo{year}{2007}),
  \urlprefix\url{http://link.aps.org/doi/10.1103/PhysRevLett.98.128301}.

\bibitem[{\citenamefont{Stellbrink et~al.}(2000)\citenamefont{Stellbrink,
  Allgaier, Monkenbusch, Richter, Lang, Likos, Watzlawek, L{\"o}wen, Ehlers,
  and Schleger}}]{stel:00}
\bibinfo{author}{\bibfnamefont{J.}~\bibnamefont{Stellbrink}},
  \bibinfo{author}{\bibfnamefont{J.}~\bibnamefont{Allgaier}},
  \bibinfo{author}{\bibfnamefont{M.}~\bibnamefont{Monkenbusch}},
  \bibinfo{author}{\bibfnamefont{D.}~\bibnamefont{Richter}},
  \bibinfo{author}{\bibfnamefont{A.}~\bibnamefont{Lang}},
  \bibinfo{author}{\bibfnamefont{C.}~\bibnamefont{Likos}},
  \bibinfo{author}{\bibfnamefont{M.}~\bibnamefont{Watzlawek}},
  \bibinfo{author}{\bibfnamefont{H.}~\bibnamefont{L{\"o}wen}},
  \bibinfo{author}{\bibfnamefont{G.}~\bibnamefont{Ehlers}}, \bibnamefont{and}
  \bibinfo{author}{\bibfnamefont{P.}~\bibnamefont{Schleger}}, in
  \emph{\bibinfo{booktitle}{Trends in Colloid and Interface Science XIV}},
  edited by \bibinfo{editor}{\bibfnamefont{V.}~\bibnamefont{Buckin}}
  (\bibinfo{publisher}{Springer Berlin Heidelberg}, \bibinfo{year}{2000}), vol.
  \bibinfo{volume}{115} of \emph{\bibinfo{series}{Progress in Colloid and
  Polymer Science}}, pp. \bibinfo{pages}{88--92}.

\bibitem[{\citenamefont{Grest et~al.}(1987)\citenamefont{Grest, Kremer, and
  Witten}}]{gres:87}
\bibinfo{author}{\bibfnamefont{G.~S.} \bibnamefont{Grest}},
  \bibinfo{author}{\bibfnamefont{K.}~\bibnamefont{Kremer}}, \bibnamefont{and}
  \bibinfo{author}{\bibfnamefont{T.~A.} \bibnamefont{Witten}},
  \bibinfo{journal}{Macromolecules} \textbf{\bibinfo{volume}{20}},
  \bibinfo{pages}{1376} (\bibinfo{year}{1987}).

\bibitem[{\citenamefont{Grest et~al.}(1989)\citenamefont{Grest, Kremer, Milner,
  and Witten}}]{gres:89}
\bibinfo{author}{\bibfnamefont{G.~S.} \bibnamefont{Grest}},
  \bibinfo{author}{\bibfnamefont{K.}~\bibnamefont{Kremer}},
  \bibinfo{author}{\bibfnamefont{S.~T.} \bibnamefont{Milner}},
  \bibnamefont{and} \bibinfo{author}{\bibfnamefont{T.~A.}
  \bibnamefont{Witten}}, \bibinfo{journal}{Macromolecules}
  \textbf{\bibinfo{volume}{22}}, \bibinfo{pages}{1904} (\bibinfo{year}{1989}).

\bibitem[{\citenamefont{Ripoll et~al.}(2006{\natexlab{a}})\citenamefont{Ripoll,
  Winkler, and Gompper}}]{Ripoll_PRL_2006}
\bibinfo{author}{\bibfnamefont{M.}~\bibnamefont{Ripoll}},
  \bibinfo{author}{\bibfnamefont{R.~G.} \bibnamefont{Winkler}},
  \bibnamefont{and} \bibinfo{author}{\bibfnamefont{G.}~\bibnamefont{Gompper}},
  \bibinfo{journal}{Phys. Rev. Lett.} \textbf{\bibinfo{volume}{96}},
  \bibinfo{pages}{188302} (\bibinfo{year}{2006}{\natexlab{a}}).

\bibitem[{\citenamefont{Singh et~al.}(2012)\citenamefont{Singh, Fedosov,
  Chatterji, Winkler, and Gompper}}]{sing:12}
\bibinfo{author}{\bibfnamefont{S.~P.} \bibnamefont{Singh}},
  \bibinfo{author}{\bibfnamefont{D.~A.} \bibnamefont{Fedosov}},
  \bibinfo{author}{\bibfnamefont{A.}~\bibnamefont{Chatterji}},
  \bibinfo{author}{\bibfnamefont{R.~G.} \bibnamefont{Winkler}},
  \bibnamefont{and} \bibinfo{author}{\bibfnamefont{G.}~\bibnamefont{Gompper}},
  \bibinfo{journal}{J. Phys.: Condens. Matter} \textbf{\bibinfo{volume}{24}},
  \bibinfo{pages}{464103} (\bibinfo{year}{2012}).

\bibitem[{\citenamefont{Singh et~al.}(2011)\citenamefont{Singh, Winkler, and
  Gompper}}]{sing:11}
\bibinfo{author}{\bibfnamefont{S.~P.} \bibnamefont{Singh}},
  \bibinfo{author}{\bibfnamefont{R.~G.} \bibnamefont{Winkler}},
  \bibnamefont{and} \bibinfo{author}{\bibfnamefont{G.}~\bibnamefont{Gompper}},
  \bibinfo{journal}{Phys. Rev. Lett.} \textbf{\bibinfo{volume}{107}},
  \bibinfo{pages}{158301} (\bibinfo{year}{2011}).

\bibitem[{\citenamefont{Singh et~al.}(2013)\citenamefont{Singh, Chatterji,
  Gompper, and Winkler}}]{Singh_MACRO_2013}
\bibinfo{author}{\bibfnamefont{S.~P.} \bibnamefont{Singh}},
  \bibinfo{author}{\bibfnamefont{A.}~\bibnamefont{Chatterji}},
  \bibinfo{author}{\bibfnamefont{G.}~\bibnamefont{Gompper}}, \bibnamefont{and}
  \bibinfo{author}{\bibfnamefont{R.~G.} \bibnamefont{Winkler}},
  \bibinfo{journal}{Macromolecules} \textbf{\bibinfo{volume}{46}},
  \bibinfo{pages}{8026} (\bibinfo{year}{2013}).

\bibitem[{\citenamefont{Gupta et~al.}(2012)\citenamefont{Gupta, Kundu,
  Stellbrink, Willner, Allgaier, and Richter}}]{gupt:12}
\bibinfo{author}{\bibfnamefont{S.}~\bibnamefont{Gupta}},
  \bibinfo{author}{\bibfnamefont{S.}~\bibnamefont{Kundu}},
  \bibinfo{author}{\bibfnamefont{J.}~\bibnamefont{Stellbrink}},
  \bibinfo{author}{\bibfnamefont{L.}~\bibnamefont{Willner}},
  \bibinfo{author}{\bibfnamefont{J.}~\bibnamefont{Allgaier}}, \bibnamefont{and}
  \bibinfo{author}{\bibfnamefont{D.}~\bibnamefont{Richter}},
  \bibinfo{journal}{J. Phys: Condens. Matter} \textbf{\bibinfo{volume}{24}},
  \bibinfo{pages}{464102} (\bibinfo{year}{2012}).

\bibitem[{\citenamefont{Sablic et~al.}(2017)\citenamefont{Sablic, Praprotnik,
  and Delgado-Buscalioni}}]{sabl:17}
\bibinfo{author}{\bibfnamefont{J.}~\bibnamefont{Sablic}},
  \bibinfo{author}{\bibfnamefont{M.}~\bibnamefont{Praprotnik}},
  \bibnamefont{and}
  \bibinfo{author}{\bibfnamefont{R.}~\bibnamefont{Delgado-Buscalioni}},
  \bibinfo{journal}{Soft Matter} \textbf{\bibinfo{volume}{13}},
  \bibinfo{pages}{4971} (\bibinfo{year}{2017}),
  \urlprefix\url{http://dx.doi.org/10.1039/C7SM00364A}.

\bibitem[{\citenamefont{Malevanets and Kapral}(1999)}]{Malevanets_MSM_1999}
\bibinfo{author}{\bibfnamefont{A.}~\bibnamefont{Malevanets}} \bibnamefont{and}
  \bibinfo{author}{\bibfnamefont{R.}~\bibnamefont{Kapral}},
  \bibinfo{journal}{J. Chem. Phys.} \textbf{\bibinfo{volume}{110}},
  \bibinfo{pages}{8605} (\bibinfo{year}{1999}).

\bibitem[{\citenamefont{Kapral}(2008)}]{Kapral_ACP_2008}
\bibinfo{author}{\bibfnamefont{R.}~\bibnamefont{Kapral}},
  \bibinfo{journal}{Adv. Chem. Phys.} \textbf{\bibinfo{volume}{140}},
  \bibinfo{pages}{89} (\bibinfo{year}{2008}).

\bibitem[{\citenamefont{Gompper et~al.}(2009)\citenamefont{Gompper, Ihle,
  Kroll, and Winkler}}]{Gompper_APS_2009}
\bibinfo{author}{\bibfnamefont{G.}~\bibnamefont{Gompper}},
  \bibinfo{author}{\bibfnamefont{T.}~\bibnamefont{Ihle}},
  \bibinfo{author}{\bibfnamefont{D.~M.} \bibnamefont{Kroll}}, \bibnamefont{and}
  \bibinfo{author}{\bibfnamefont{R.~G.} \bibnamefont{Winkler}},
  \bibinfo{journal}{Adv. Polym. Sci.} \textbf{\bibinfo{volume}{221}},
  \bibinfo{pages}{1} (\bibinfo{year}{2009}).

\bibitem[{\citenamefont{Malevanets and Yeomans}(2000)}]{male00}
\bibinfo{author}{\bibfnamefont{A.}~\bibnamefont{Malevanets}} \bibnamefont{and}
  \bibinfo{author}{\bibfnamefont{J.~M.} \bibnamefont{Yeomans}},
  \bibinfo{journal}{Europhys. Lett.} \textbf{\bibinfo{volume}{52}},
  \bibinfo{pages}{231} (\bibinfo{year}{2000}).

\bibitem[{\citenamefont{Huang et~al.}(2012{\natexlab{a}})\citenamefont{Huang,
  Gompper, and Winkler}}]{huan:12}
\bibinfo{author}{\bibfnamefont{C.-C.} \bibnamefont{Huang}},
  \bibinfo{author}{\bibfnamefont{G.}~\bibnamefont{Gompper}}, \bibnamefont{and}
  \bibinfo{author}{\bibfnamefont{R.~G.} \bibnamefont{Winkler}},
  \bibinfo{journal}{J. Phys.: Condens. Matter} \textbf{\bibinfo{volume}{24}},
  \bibinfo{pages}{284131} (\bibinfo{year}{2012}{\natexlab{a}}).

\bibitem[{\citenamefont{Huang et~al.}(2013)\citenamefont{Huang, Gompper, and
  Winkler}}]{huan:13}
\bibinfo{author}{\bibfnamefont{C.~C.} \bibnamefont{Huang}},
  \bibinfo{author}{\bibfnamefont{G.}~\bibnamefont{Gompper}}, \bibnamefont{and}
  \bibinfo{author}{\bibfnamefont{R.~G.} \bibnamefont{Winkler}},
  \bibinfo{journal}{J. Chem. Phys.} \textbf{\bibinfo{volume}{138}},
  \bibinfo{pages}{144902} (\bibinfo{year}{2013}).

\bibitem[{\citenamefont{Kikuchi et~al.}(2002)\citenamefont{Kikuchi, Gent, and
  Yeomans}}]{kiku:02}
\bibinfo{author}{\bibfnamefont{N.}~\bibnamefont{Kikuchi}},
  \bibinfo{author}{\bibfnamefont{A.}~\bibnamefont{Gent}}, \bibnamefont{and}
  \bibinfo{author}{\bibfnamefont{J.~M.} \bibnamefont{Yeomans}},
  \bibinfo{journal}{Eur. Phys. J. E} \textbf{\bibinfo{volume}{9}},
  \bibinfo{pages}{63} (\bibinfo{year}{2002}),
  \urlprefix\url{http://dx.doi.org/10.1140/epje/i2002-10056-6}.

\bibitem[{\citenamefont{Webster and Yeomans}(2005)}]{webs:05}
\bibinfo{author}{\bibfnamefont{M.~A.} \bibnamefont{Webster}} \bibnamefont{and}
  \bibinfo{author}{\bibfnamefont{J.~M.} \bibnamefont{Yeomans}},
  \bibinfo{journal}{J. Chem. Phys.} \textbf{\bibinfo{volume}{122}},
  \bibinfo{pages}{164903} (\bibinfo{year}{2005}).

\bibitem[{\citenamefont{Ryder and Yeomans}(2006)}]{ryde:06}
\bibinfo{author}{\bibfnamefont{J.~F.} \bibnamefont{Ryder}} \bibnamefont{and}
  \bibinfo{author}{\bibfnamefont{J.~M.} \bibnamefont{Yeomans}},
  \bibinfo{journal}{J. Chem. Phys.} \textbf{\bibinfo{volume}{125}},
  \bibinfo{pages}{194906} (\bibinfo{year}{2006}).

\bibitem[{\citenamefont{Frank and Winkler}(2008)}]{fran:08}
\bibinfo{author}{\bibfnamefont{S.}~\bibnamefont{Frank}} \bibnamefont{and}
  \bibinfo{author}{\bibfnamefont{R.~G.} \bibnamefont{Winkler}},
  \bibinfo{journal}{EPL} \textbf{\bibinfo{volume}{83}}, \bibinfo{pages}{38004}
  (\bibinfo{year}{2008}).

\bibitem[{\citenamefont{Frank and Winkler}(2009)}]{fran:09}
\bibinfo{author}{\bibfnamefont{S.}~\bibnamefont{Frank}} \bibnamefont{and}
  \bibinfo{author}{\bibfnamefont{R.~G.} \bibnamefont{Winkler}},
  \bibinfo{journal}{J. Chem. Phys.} \textbf{\bibinfo{volume}{131}},
  \bibinfo{pages}{234905} (\bibinfo{year}{2009}).

\bibitem[{\citenamefont{Huang et~al.}(2010{\natexlab{a}})\citenamefont{Huang,
  Winkler, Sutmann, and Gompper}}]{Huang_MAM_2010}
\bibinfo{author}{\bibfnamefont{C.-C.} \bibnamefont{Huang}},
  \bibinfo{author}{\bibfnamefont{R.~G.} \bibnamefont{Winkler}},
  \bibinfo{author}{\bibfnamefont{G.}~\bibnamefont{Sutmann}}, \bibnamefont{and}
  \bibinfo{author}{\bibfnamefont{G.}~\bibnamefont{Gompper}},
  \bibinfo{journal}{Macromolecules} \textbf{\bibinfo{volume}{43}},
  \bibinfo{pages}{10107} (\bibinfo{year}{2010}{\natexlab{a}}).

\bibitem[{\citenamefont{Padding and Louis}(2004)}]{padd:04}
\bibinfo{author}{\bibfnamefont{J.~T.} \bibnamefont{Padding}} \bibnamefont{and}
  \bibinfo{author}{\bibfnamefont{A.~A.} \bibnamefont{Louis}},
  \bibinfo{journal}{Phys. Rev. Lett.} \textbf{\bibinfo{volume}{93}},
  \bibinfo{pages}{220601} (\bibinfo{year}{2004}).

\bibitem[{\citenamefont{Ripoll et~al.}(2006{\natexlab{b}})\citenamefont{Ripoll,
  Winkler, and Gompper}}]{ripo:06}
\bibinfo{author}{\bibfnamefont{M.}~\bibnamefont{Ripoll}},
  \bibinfo{author}{\bibfnamefont{R.~G.} \bibnamefont{Winkler}},
  \bibnamefont{and} \bibinfo{author}{\bibfnamefont{G.}~\bibnamefont{Gompper}},
  \bibinfo{journal}{Phys. Rev. Lett.} \textbf{\bibinfo{volume}{96}},
  \bibinfo{pages}{188302} (\bibinfo{year}{2006}{\natexlab{b}}).

\bibitem[{\citenamefont{Fedosov et~al.}(2012)\citenamefont{Fedosov, Singh,
  Chatterji, Winkler, and Gompper}}]{fedo:12}
\bibinfo{author}{\bibfnamefont{D.~A.} \bibnamefont{Fedosov}},
  \bibinfo{author}{\bibfnamefont{S.~P.} \bibnamefont{Singh}},
  \bibinfo{author}{\bibfnamefont{A.}~\bibnamefont{Chatterji}},
  \bibinfo{author}{\bibfnamefont{R.~G.} \bibnamefont{Winkler}},
  \bibnamefont{and} \bibinfo{author}{\bibfnamefont{G.}~\bibnamefont{Gompper}},
  \bibinfo{journal}{Soft Matter} \textbf{\bibinfo{volume}{8}},
  \bibinfo{pages}{4109} (\bibinfo{year}{2012}).

\bibitem[{\citenamefont{Nikoubashman and Likos}(2010{\natexlab{b}})}]{niko:10}
\bibinfo{author}{\bibfnamefont{A.}~\bibnamefont{Nikoubashman}}
  \bibnamefont{and} \bibinfo{author}{\bibfnamefont{C.~N.} \bibnamefont{Likos}},
  \bibinfo{journal}{J. Chem. Phys.} \textbf{\bibinfo{volume}{133}},
  \bibinfo{pages}{074901} (\bibinfo{year}{2010}{\natexlab{b}}).

\bibitem[{\citenamefont{Mcwhirter et~al.}(2009)\citenamefont{Mcwhirter,
  Noguchi, and Gompper}}]{mcwh:09}
\bibinfo{author}{\bibfnamefont{J.~L.} \bibnamefont{Mcwhirter}},
  \bibinfo{author}{\bibfnamefont{H.}~\bibnamefont{Noguchi}}, \bibnamefont{and}
  \bibinfo{author}{\bibfnamefont{G.}~\bibnamefont{Gompper}},
  \bibinfo{journal}{Proc. Natl. Acad. Sci. USA} \textbf{\bibinfo{volume}{106}},
  \bibinfo{pages}{6039} (\bibinfo{year}{2009}).

\bibitem[{\citenamefont{Tao and Kapral}(2010)}]{tao:10}
\bibinfo{author}{\bibfnamefont{Y.-G.} \bibnamefont{Tao}} \bibnamefont{and}
  \bibinfo{author}{\bibfnamefont{R.}~\bibnamefont{Kapral}},
  \bibinfo{journal}{Soft Matter} \textbf{\bibinfo{volume}{6}},
  \bibinfo{pages}{756} (\bibinfo{year}{2010}).

\bibitem[{\citenamefont{Z{\"o}ttl and Stark}(2012)}]{zoet:12}
\bibinfo{author}{\bibfnamefont{A.}~\bibnamefont{Z{\"o}ttl}} \bibnamefont{and}
  \bibinfo{author}{\bibfnamefont{H.}~\bibnamefont{Stark}},
  \bibinfo{journal}{Phys. Rev. Lett.} \textbf{\bibinfo{volume}{108}},
  \bibinfo{pages}{218104} (\bibinfo{year}{2012}).

\bibitem[{\citenamefont{Elgeti and Gompper}(2013)}]{elge:13}
\bibinfo{author}{\bibfnamefont{J.}~\bibnamefont{Elgeti}} \bibnamefont{and}
  \bibinfo{author}{\bibfnamefont{G.}~\bibnamefont{Gompper}},
  \bibinfo{journal}{Proc. Natl. Acad. Sci. USA} \textbf{\bibinfo{volume}{110}},
  \bibinfo{pages}{4470} (\bibinfo{year}{2013}).

\bibitem[{\citenamefont{Reigh et~al.}(2012)\citenamefont{Reigh, Winkler, and
  Gompper}}]{reig:12}
\bibinfo{author}{\bibfnamefont{S.~Y.} \bibnamefont{Reigh}},
  \bibinfo{author}{\bibfnamefont{R.~G.} \bibnamefont{Winkler}},
  \bibnamefont{and} \bibinfo{author}{\bibfnamefont{G.}~\bibnamefont{Gompper}},
  \bibinfo{journal}{Soft Matter} \textbf{\bibinfo{volume}{8}},
  \bibinfo{pages}{4363} (\bibinfo{year}{2012}).

\bibitem[{\citenamefont{Theers and Winkler}(2014)}]{thee:14}
\bibinfo{author}{\bibfnamefont{M.}~\bibnamefont{Theers}} \bibnamefont{and}
  \bibinfo{author}{\bibfnamefont{R.~G.} \bibnamefont{Winkler}},
  \bibinfo{journal}{Soft Matter} \textbf{\bibinfo{volume}{10}},
  \bibinfo{pages}{5894} (\bibinfo{year}{2014}),
  \urlprefix\url{http://dx.doi.org/10.1039/C4SM00770K}.

\bibitem[{\citenamefont{Theers et~al.}(2016)\citenamefont{Theers, Westphal,
  Gompper, and Winkler}}]{thee:16.1}
\bibinfo{author}{\bibfnamefont{M.}~\bibnamefont{Theers}},
  \bibinfo{author}{\bibfnamefont{E.}~\bibnamefont{Westphal}},
  \bibinfo{author}{\bibfnamefont{G.}~\bibnamefont{Gompper}}, \bibnamefont{and}
  \bibinfo{author}{\bibfnamefont{R.~G.} \bibnamefont{Winkler}},
  \bibinfo{journal}{Soft Matter} \textbf{\bibinfo{volume}{12}},
  \bibinfo{pages}{7372} (\bibinfo{year}{2016}),
  \urlprefix\url{http://dx.doi.org/10.1039/C6SM01424K}.

\bibitem[{\citenamefont{Yang et~al.}(2014)\citenamefont{Yang, Wysocki, and
  Ripoll}}]{yang:14.1}
\bibinfo{author}{\bibfnamefont{M.}~\bibnamefont{Yang}},
  \bibinfo{author}{\bibfnamefont{A.}~\bibnamefont{Wysocki}}, \bibnamefont{and}
  \bibinfo{author}{\bibfnamefont{M.}~\bibnamefont{Ripoll}},
  \bibinfo{journal}{Soft Matter} \textbf{\bibinfo{volume}{10}},
  \bibinfo{pages}{6208} (\bibinfo{year}{2014}),
  \urlprefix\url{http://dx.doi.org/10.1039/C4SM00621F}.

\bibitem[{\citenamefont{Hu et~al.}(2015)\citenamefont{Hu, Wysocki, Winkler, and
  Gompper}}]{hu:15}
\bibinfo{author}{\bibfnamefont{J.}~\bibnamefont{Hu}},
  \bibinfo{author}{\bibfnamefont{A.}~\bibnamefont{Wysocki}},
  \bibinfo{author}{\bibfnamefont{R.~G.} \bibnamefont{Winkler}},
  \bibnamefont{and} \bibinfo{author}{\bibfnamefont{G.}~\bibnamefont{Gompper}},
  \bibinfo{journal}{Sci. Rep.} \textbf{\bibinfo{volume}{5}},
  \bibinfo{pages}{9586} (\bibinfo{year}{2015}),
  \urlprefix\url{http://dx.doi.org/10.1038/srep09586}.

\bibitem[{\citenamefont{Allen and Tildesley}(1987)}]{allen87}
\bibinfo{author}{\bibfnamefont{M.~P.} \bibnamefont{Allen}} \bibnamefont{and}
  \bibinfo{author}{\bibfnamefont{D.~J.} \bibnamefont{Tildesley}},
  \emph{\bibinfo{title}{Computer Simulation of Liquids}}
  (\bibinfo{publisher}{Clarendon Press}, \bibinfo{address}{Oxford},
  \bibinfo{year}{1987}).

\bibitem[{\citenamefont{Huang et~al.}(2012{\natexlab{b}})\citenamefont{Huang,
  Gompper, and Winkler}}]{huan:12.1}
\bibinfo{author}{\bibfnamefont{C.-C.} \bibnamefont{Huang}},
  \bibinfo{author}{\bibfnamefont{G.}~\bibnamefont{Gompper}}, \bibnamefont{and}
  \bibinfo{author}{\bibfnamefont{R.~G.} \bibnamefont{Winkler}},
  \bibinfo{journal}{Phys. Rev. E} \textbf{\bibinfo{volume}{86}},
  \bibinfo{pages}{056711} (\bibinfo{year}{2012}{\natexlab{b}}).

\bibitem[{\citenamefont{Mussawisade et~al.}(2005)\citenamefont{Mussawisade,
  Ripoll, Winkler, and Gompper}}]{muss:05}
\bibinfo{author}{\bibfnamefont{K.}~\bibnamefont{Mussawisade}},
  \bibinfo{author}{\bibfnamefont{M.}~\bibnamefont{Ripoll}},
  \bibinfo{author}{\bibfnamefont{R.~G.} \bibnamefont{Winkler}},
  \bibnamefont{and} \bibinfo{author}{\bibfnamefont{G.}~\bibnamefont{Gompper}},
  \bibinfo{journal}{J. Chem. Phys.} \textbf{\bibinfo{volume}{123}},
  \bibinfo{pages}{144905} (\bibinfo{year}{2005}).

\bibitem[{\citenamefont{Huang et~al.}(2010{\natexlab{b}})\citenamefont{Huang,
  Chatterji, Sutmann, Gompper, and Winkler}}]{Huang_JCPS_2010}
\bibinfo{author}{\bibfnamefont{C.-C.} \bibnamefont{Huang}},
  \bibinfo{author}{\bibfnamefont{A.}~\bibnamefont{Chatterji}},
  \bibinfo{author}{\bibfnamefont{G.}~\bibnamefont{Sutmann}},
  \bibinfo{author}{\bibfnamefont{G.}~\bibnamefont{Gompper}}, \bibnamefont{and}
  \bibinfo{author}{\bibfnamefont{R.~G.} \bibnamefont{Winkler}},
  \bibinfo{journal}{J. Comp. Phys.} \textbf{\bibinfo{volume}{229}},
  \bibinfo{pages}{168} (\bibinfo{year}{2010}{\natexlab{b}}).

\bibitem[{\citenamefont{Huang et~al.}(2015)\citenamefont{Huang, Varghese,
  Gompper, and Winkler}}]{huan:15}
\bibinfo{author}{\bibfnamefont{C.-C.} \bibnamefont{Huang}},
  \bibinfo{author}{\bibfnamefont{A.}~\bibnamefont{Varghese}},
  \bibinfo{author}{\bibfnamefont{G.}~\bibnamefont{Gompper}}, \bibnamefont{and}
  \bibinfo{author}{\bibfnamefont{R.~G.} \bibnamefont{Winkler}},
  \bibinfo{journal}{Phys. Rev. E} \textbf{\bibinfo{volume}{91}},
  \bibinfo{pages}{013310} (\bibinfo{year}{2015}),
  \urlprefix\url{http://link.aps.org/doi/10.1103/PhysRevE.91.013310}.

\bibitem[{\citenamefont{Ihle and Kroll}(2003)}]{ihle:03}
\bibinfo{author}{\bibfnamefont{T.}~\bibnamefont{Ihle}} \bibnamefont{and}
  \bibinfo{author}{\bibfnamefont{D.~M.} \bibnamefont{Kroll}},
  \bibinfo{journal}{Phys. Rev. E} \textbf{\bibinfo{volume}{67}},
  \bibinfo{pages}{066706} (\bibinfo{year}{2003}).

\bibitem[{\citenamefont{Kikuchi et~al.}(2003)\citenamefont{Kikuchi, Pooley,
  Ryder, and Yeomans}}]{kiku:03}
\bibinfo{author}{\bibfnamefont{N.}~\bibnamefont{Kikuchi}},
  \bibinfo{author}{\bibfnamefont{C.~M.} \bibnamefont{Pooley}},
  \bibinfo{author}{\bibfnamefont{J.~F.} \bibnamefont{Ryder}}, \bibnamefont{and}
  \bibinfo{author}{\bibfnamefont{J.~M.} \bibnamefont{Yeomans}},
  \bibinfo{journal}{J. Chem. Phys.} \textbf{\bibinfo{volume}{119}},
  \bibinfo{pages}{6388} (\bibinfo{year}{2003}).

\bibitem[{\citenamefont{Pooley and Yeomans}(2005)}]{pool:05}
\bibinfo{author}{\bibfnamefont{C.~M.} \bibnamefont{Pooley}} \bibnamefont{and}
  \bibinfo{author}{\bibfnamefont{J.~M.} \bibnamefont{Yeomans}},
  \bibinfo{journal}{J. Phys. Chem. B} \textbf{\bibinfo{volume}{109}},
  \bibinfo{pages}{6505} (\bibinfo{year}{2005}).

\bibitem[{\citenamefont{Noguchi and Gompper}(2008)}]{nogu:08}
\bibinfo{author}{\bibfnamefont{H.}~\bibnamefont{Noguchi}} \bibnamefont{and}
  \bibinfo{author}{\bibfnamefont{G.}~\bibnamefont{Gompper}},
  \bibinfo{journal}{Phys. Rev. E} \textbf{\bibinfo{volume}{78}},
  \bibinfo{pages}{016706} (\bibinfo{year}{2008}).

\bibitem[{\citenamefont{Winkler and Huang}(2009)}]{wink:09}
\bibinfo{author}{\bibfnamefont{R.~G.} \bibnamefont{Winkler}} \bibnamefont{and}
  \bibinfo{author}{\bibfnamefont{C.-C.} \bibnamefont{Huang}},
  \bibinfo{journal}{J. Chem. Phys.} \textbf{\bibinfo{volume}{130}},
  \bibinfo{pages}{074907} (\bibinfo{year}{2009}).

\bibitem[{\citenamefont{Westphal et~al.}(2014)\citenamefont{Westphal, Singh,
  Huang, Gompper, and Winkler}}]{west:14}
\bibinfo{author}{\bibfnamefont{E.}~\bibnamefont{Westphal}},
  \bibinfo{author}{\bibfnamefont{S.~P.} \bibnamefont{Singh}},
  \bibinfo{author}{\bibfnamefont{C.-C.} \bibnamefont{Huang}},
  \bibinfo{author}{\bibfnamefont{G.}~\bibnamefont{Gompper}}, \bibnamefont{and}
  \bibinfo{author}{\bibfnamefont{R.~G.} \bibnamefont{Winkler}},
  \bibinfo{journal}{Comput. Phys. Comm.} \textbf{\bibinfo{volume}{185}},
  \bibinfo{pages}{495} (\bibinfo{year}{2014}).

\bibitem[{\citenamefont{Daoud and Cotton}(1982)}]{daou82}
\bibinfo{author}{\bibfnamefont{M.}~\bibnamefont{Daoud}} \bibnamefont{and}
  \bibinfo{author}{\bibfnamefont{J.}~\bibnamefont{Cotton}},
  \bibinfo{journal}{Journal de Physique} \textbf{\bibinfo{volume}{43}},
  \bibinfo{pages}{531} (\bibinfo{year}{1982}).

\bibitem[{\citenamefont{Birshtein et~al.}(1986)\citenamefont{Birshtein,
  Zhulina, and Borisov}}]{birsh86}
\bibinfo{author}{\bibfnamefont{T.}~\bibnamefont{Birshtein}},
  \bibinfo{author}{\bibfnamefont{E.}~\bibnamefont{Zhulina}}, \bibnamefont{and}
  \bibinfo{author}{\bibfnamefont{O.}~\bibnamefont{Borisov}},
  \bibinfo{journal}{Polymer} \textbf{\bibinfo{volume}{27}},
  \bibinfo{pages}{1078} (\bibinfo{year}{1986}).

\bibitem[{\citenamefont{Singh et~al.}(2014)\citenamefont{Singh, Huang,
  Westphal, Gompper, and Winkler}}]{sing:14}
\bibinfo{author}{\bibfnamefont{S.~P.} \bibnamefont{Singh}},
  \bibinfo{author}{\bibfnamefont{C.-C.} \bibnamefont{Huang}},
  \bibinfo{author}{\bibfnamefont{E.}~\bibnamefont{Westphal}},
  \bibinfo{author}{\bibfnamefont{G.}~\bibnamefont{Gompper}}, \bibnamefont{and}
  \bibinfo{author}{\bibfnamefont{R.~G.} \bibnamefont{Winkler}},
  \bibinfo{journal}{J. Chem. Phys.} \textbf{\bibinfo{volume}{141}},
  \bibinfo{pages}{084901} (\bibinfo{year}{2014}),
  \urlprefix\url{http://scitation.aip.org/content/aip/journal/jcp/141/8/10.1063/1.4893766}.

\end{thebibliography}

\end{document}